\documentclass[leqno, 12pt]{article}

\usepackage{abstract,amssymb,amsmath,amsfonts,booktabs,enumitem,eurosym, ragged2e, geometry,ulem,graphicx,caption,color,setspace,sectsty,comment,footmisc,footnote,caption,lipsum,mathtools,pdflscape,subfigure,array,makecell,csquotes, fancyhdr, setspace}
\usepackage{indentfirst}
\usepackage{threeparttablex}
\usepackage{longtable}
\usepackage{booktabs}
\usepackage{makecell}

\usepackage[authordate,backend=bibtex]{biblatex-chicago}
\DeclareFieldFormat[article]{title}{\mkbibquote{#1}} 
\DeclareFieldFormat[thesis]{title}{\mkbibemph{#1}} 
\usepackage[toc,page]{appendix}

\usepackage{hyperref}
\hypersetup{
    colorlinks=true,
    linkcolor=blue,
    filecolor=magenta,      
    urlcolor=blue,
    citecolor=blue
}

\newtheorem{theorem}{Theorem}

\newtheorem{proposition}{Proposition}
\newtheorem{lemma}[theorem]{Lemma}
\newtheorem{subprop}{Proposition}[proposition] 

\newenvironment{proof}[1][Proof]{\noindent\textbf{#1.} }{\ \rule{0.5em}{0.5em}}

\AtEndEnvironment{subequations}{\setcounter{mysub}{0}}

\newcounter{mysub}
\newcolumntype{L}[1]{>{\raggedright\let\newline\\arraybackslash\hspace{0pt}}m{#1}}
\newcolumntype{C}[1]{>{\centering\let\newline\\arraybackslash\hspace{0pt}}m{#1}}
\newcolumntype{R}[1]{>{\raggedleft\let\newline\\arraybackslash\hspace{0pt}}m{#1}}

\makesavenoteenv{tabular}



\begin{document}

\begin{titlepage}
\title{\vspace{-1.9cm}Seeing Through Color Blindness:\\Social Networks as a Mechanism for Discrimination\thanks{I extend deep thanks to Kenneth Arrow, Ben Golub, Lawrence Katz, Abraham Wickelgren, and---above all---to God.  I also thank Ian Ayres, Raj Chetty, Krishna Dasaratha, Ed Glaeser, Louis Kaplow, Yair Listokin, Chang Liu, Sriniketh Nagavarapu, Nathan Nunn, Chinwuba Okafor, Nkasi Okafor, Geoffrey Rothwell, Kathryn Spier, and Ebonya Washington for helpful comments.  This work has benefited from thoughtful feedback at the Law and Economics Seminar at Harvard Law School.  I also give thanks for outstanding research assistance from Montse Trujillo.  This research was supported by a Ford Foundation Predoctoral Fellowship administered by the National Academies of Sciences, Engineering, and Medicine; the National Institute on Aging, Grant Number T32-AG000186; and the Radcliffe Institute for Advanced Study at Harvard University.  An anonymous referee and editor Richard Holden provided very valuable suggestions.}}

\author{Chika O. Okafor\thanks{Okafor: Northwestern University Pritzker School of Law, 375 East Chicago Avenue, Chicago, IL 60611.}
\\\textit{Northwestern University}
}
%
\date{}
\maketitle

\begin{center}
\href{https://doi.org/10.1086/735081}{CLICK HERE FOR PUBLICATION VERSION}
\rule{0.85\textwidth}{0.4pt}\\[0.4em]
\begin{minipage}{0.85\textwidth}\small\singlespacing\noindent
Published as: Chika O. Okafor, ``Seeing Through
Color Blindness: Social Networks as a Mechanism for Discrimination,''
\emph{Journal of Law and Economics} 68, no.~3 (2025): 519--560,
\href{https://doi.org/10.1086/735081}{https://doi.org/10.1086/735081}.
\copyright{} 2025 by The University of Chicago. All rights reserved.
Please cite the published version. Posted under a CC BY-NC-ND 4.0 license.\\
\end{minipage}\\[0.4em]
\rule{0.85\textwidth}{0.4pt}
\end{center}

\begin{abstract}
\noindent I study labor markets in which firms both hire via referrals and are race blind or color-blind. I develop an employment model showing that despite initial equality in ability, employment, wages, and network structure, minorities receive disproportionately fewer jobs through referrals and lower expected wages, simply because their social group is smaller. This discriminatory outcome, which I term “social network discrimination,” arises from homophily and falls outside the dominant economics discrimination models, which are taste based and statistical. I calibrate the model using a nationally representative sample of youth networks to estimate the lower bound welfare gap caused by social network discrimination, which also disadvantages black workers. This paper isolates a potential underlying mechanism for inequality, adding to the understanding of labor-market disparities that have been widely studied across the social sciences. In doing so, the paper disproves the proposition that color-blind policies inherently promote individual merit.\\
\\
\noindent\textbf{Keywords:} discrimination, diversity, employment, homophily, inequality, labor market, meritocracy; social networks, race, affirmative action

\vspace{0in}
\end{abstract}
\end{titlepage}

\thispagestyle{empty}  

\newpage

\doublespacing

\section{Introduction} \label{sec:introduction}

In 2023, the Supreme Court upended the longstanding policy that allowed race-based affirmative action in college admissions (in \textit{Students for Fair Admissions, Inc. v. President \& Fellows of Harvard College} (600 U.S. 181 [2023])).  Its decision might threaten future efforts to achieve racial diversity on college campuses around the nation, and may eventually catalyze retrenchment of racial diversity practices in other settings like private hiring, promotions, and the government.

Conservatives have long argued that race-neutral or ``colorblind'' policies promote individual merit.\footnote{\textcite{anderson2010} lists arguments used in favor of colorblind policies: ``[s]ome advocates of color-blind policy reject racial preferences for their purportedly bad consequences: they are divisive, stigmatizing, and inefficient, depress motivation, and harm their intended beneficiaries. \textit{These advocates also claim that if only people would stop intentionally discriminating by race, then racial discrimination would end.} Other advocates of color-blind policy reject racial preferences on principle. They claim that formally race-conscious preferences and sometimes also underlying race-conscious purposes...\textit{violate the principle of merit}...'' (emphasis added).  Furthermore, \textcite{coate1993will} develops an economic model to make the claim that affirmative action can result in employers \textit{correctly} viewing the group benefiting from affirmative action as being less productive (due to the group being disincentivized, under certain conditions, to invest in their own human capital).}  But new evidence from this paper challenges that assumption, revealing that all else being equal (for example, equal education, employment, income), \textit{with} ``colorblind'' policies, members of the minority group receive fewer economic and social opportunities simply because their social group is smaller.

In this paper, I develop an employment model and assume equal ability, equal initial employment, and equal network structure between majority and minority workers.  Workers are observationally equivalent, so employers cannot determine the race or ability of prospective job candidates.\footnote{Employers do learn the ability of current employees after the initial period---based on their productivity---though employers still cannot ascertain whether they belong to the majority or minority group.  Hence, employers are fully race-blind and must use ``colorblind'' hiring policies.}  Majority and minority workers have (1) an equal chance of having a social tie (that is, an equivalent network density) and (2) an equal bias in favor of forming social ties with the same majority or minority type (that is, an equivalent type in-group bias).  Despite these equalizing assumptions, I find that the probability of a firm offering a job through referral to minority workers is \textit{lower} than their share of the labor force.  For minority workers to have a proportional chance of receiving job offers through referral, they must compensate with a stronger network density and/or type in-group bias.  The estimated welfare gap increases in a convex way with the majority group share of the labor force; in other words, the disadvantage of minority workers grows at a rate disproportionate to their minority status in the labor market.  Finally, this paper calibrates the model using the National Longitudinal Study of Adolescent to Adult Health---a nationally representative sample of youth peer networks---estimating a lower bound welfare gap between black and white workers due to what I term \textit{social network discrimination}.\par

\begin{figure}
\captionsetup{justification=centering}
\caption{Sample Illustration of Social Network Discrimination}
\centering
\includegraphics[width=.65\textwidth]{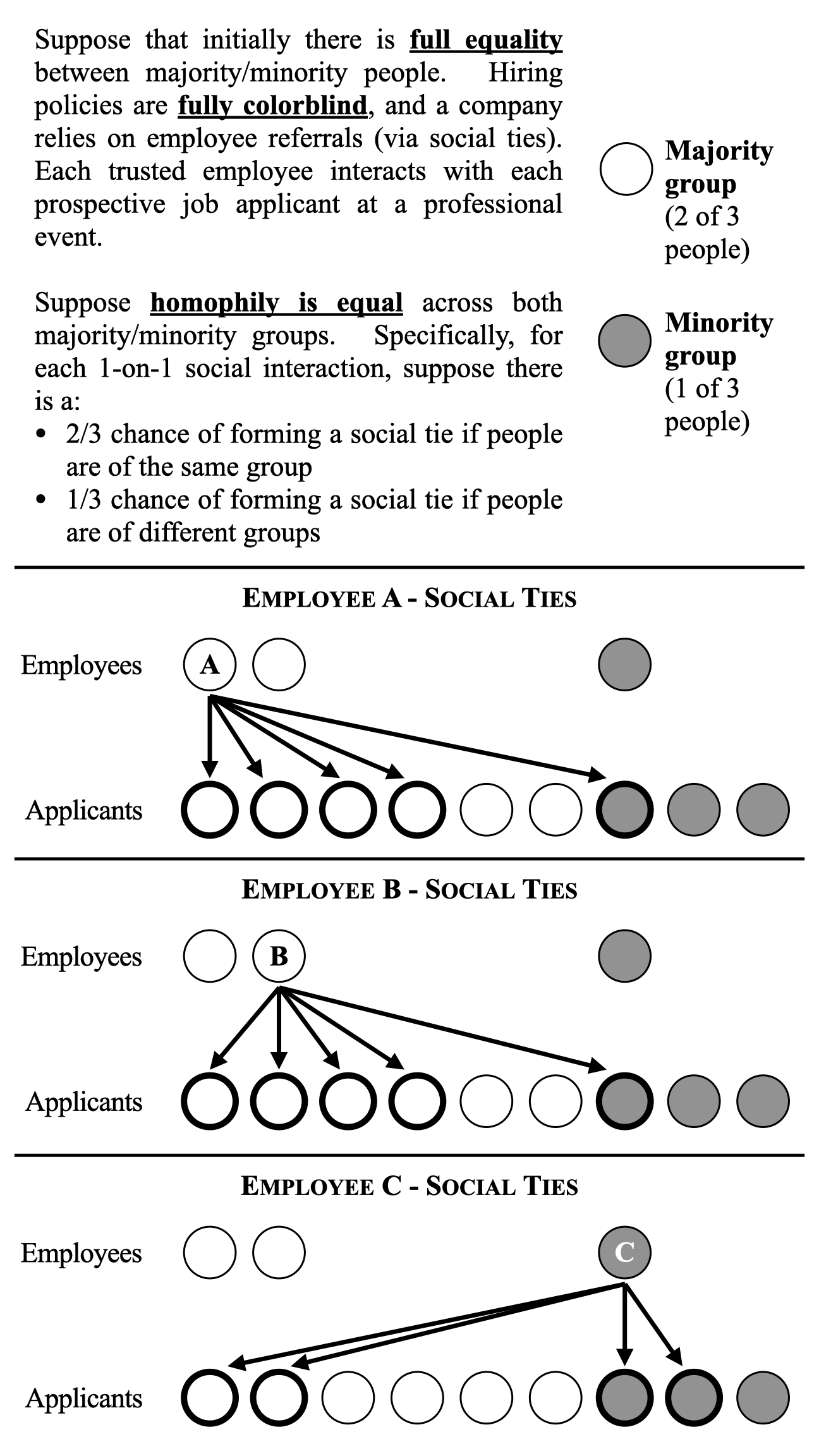} \\
\justify \footnotesize \textit{Note:} In this example, despite initial equality between the majority and minority groups, 4 out of 14 social ties are formed with minorities (or less than 29\%).  Yet minorities are 1 out of 3 people (or more than 33\%). This disproportionately low number of social ties formed with minorities is one illustration of social network discrimination. 
\label{fig:intuition}
\end{figure}%

\textit{Social network discrimination} is a term this paper introduces to the economics literature that captures the phenomenon in which minorities suffer disadvantages in economic and social opportunities, all else equal, simply because their social group is smaller.\footnote{A video explaining social network discrimination can be found online at \textcite{okaforyoutube}.}  Figure \ref{fig:intuition} depicts a simplified example that illustrates social network discrimination.  In this example, there is full initial equality between majority and minority groups---including equal homophily in the formation of social ties across groups---and fully colorblind hiring policies.\footnote{More details on the key elements involved in Figure \ref{fig:intuition} are as follows:
\begin{enumerate}[label=\alph*)]
    \item There are three trusted employees who interact with prospective job applicants at a professional event.
    \item There is full equality between majority and minority groups.  In other words, not only is employment for both majority and minority groups proportional to their share of the population, but also there is equal average qualifications and equal average ability.
    \item There is full colorblindness in the hiring policies the company uses.  In other words, the company does not use race in making their hiring decisions. \item The company relies on referrals (a common practice in the United States), which means social networks matter and employees can refer others for jobs.  Firms’ preference for referred workers is advantageous; referrals reduce the uncertainty of a job candidate’s ability.  \textcite{simon1992} mentions that ``objective'' criteria like education and work experience imperfectly communicate a candidate’s true ability to perform a particular job role.  Thus, job matching theories stress that objective criteria do not provide complete information \parencite{barron1985employer, barron1985extensive, barron1989job}.  Hiring referred employees helps compensate for this shortcoming by providing an additional subjective assessment of a candidate’s ``fit'' for the job.  Also, because a worker’s reputation with their employer is at stake, they have a vested interest in only referring well-qualified acquaintances \parencite{montgomery1991}.  Past research has estimated that a significant number of jobs in the U.S. are filled via referrals, with some estimating it at about half \parencite[See, for example,][]{holzer1987hiring, Marsden1990recruitment, montgomery1991, bewley1999wages, flynn}.
    \item Each trusted employee interacts with the prospective job applicants and makes referrals based on the social ties they form.  To understand how these social ties might form, it is important to understand the concept of \textit{homophily}.  Homophily is a widely-studied social phenomenon that captures the fact that ``birds of a feather flock together''---in other words, people are more likely to form social ties with other people with whom they share more characteristics \parencite{mcpherson2001}.
    \item Homophily is equal across both the majority and minority groups.  Specifically, for each 1-on-1 social interaction, suppose there is: (a) a 2/3 chance of forming a social tie if people are of the same majority/minority group; and (b) a 1/3 chance of forming a social tie if people are of different groups. \item The majority group is comprised of 2 out of 3 people, while the minority group is comprised of 1 out of 3 people.  Specifically, among employees, there are two majority employees and one minority employee.  Among prospective applicants, there are six majority applicants and three minority applicants.
    \end{enumerate}}
Given initial equality with equal homophily, we can see from Figure \ref{fig:intuition} that Employee A---who belongs to the majority group---will form four social ties with majority applicants and one social tie with minority applicants.  Employee B, also of the majority group, will similarly form four social ties with majority applicants and one social tie with minority applicants.  Employee C (of the minority group) will form two social ties with majority applicants and two social ties with minority applicants.  Thus, after the professional event, there are four social ties with minority applicants out of a total of 14 social ties with applicants, which means less than 29\% of social ties go to the minority group---and, consequently, less than 29\% of referrals go to minorities.  Yet minorities are over 33\% of the population (since they are 1 out of 3 people).  So, as Figure \ref{fig:intuition} shows, there will be disproportionately fewer social ties for minority applicants, despite both groups starting off equal and the company using fully colorblind hiring policies.  This disadvantage for minorities in forming social ties represents one illustration of social network discrimination.\footnote{Though this example examined an employment context, social network discrimination could arise in other settings in which economic and social opportunities are distributed based on social networks.  For example, on college campuses, key educational and career information is sometimes distributed via informal information channels linked to students' social networks.  So, social network discrimination could apply to students from any given Classroom A interacting with students from any given Classroom B---or, more generally, to any two groups of students interacting on a college campus.}

This paper formalizes social network discrimination mathematically using economic theory and begins to explore its implications on the law.  While the theory is developed for arbitrary group assignments that affect network formation, sociology suggests that race and ethnicity create the greatest divide socially \parencite{mcpherson2001}.\footnote{According to \textcite{mcpherson2001}, ``after race and ethnicity, the greatest social divides are created by age, religion, education, occupation, and gender, in approximately that order.''}  Hence, this paper's findings are relevant to debates surrounding the merits of race-conscious policies.  Many who are critical of such policies share a common assumption: in a world without historical discrimination---without Jim Crow or implicit biases---fully race-blind policies would be the best policy \parencite[see, for example,][]{anderson2010}.  Race-blind policies would yield the most meritocratic outcome, with opportunities distributed according to corresponding ability or ``merit.''\footnote{Members of the Supreme Court have echoed related sentiments, with one famous quote by Justice John Roberts stating ``[t]he way to stop discrimination on the basis of race is to stop discriminating on the basis of race'' (in \textit{Parents Involved in Community Schools v. Seattle School District No. 1} (551 U.S. 701 [2007])).  For economic analysis of colorblind policies and affirmative action that contrast with this sentiment, see, for example, \textcite{chan2003does, ray2010remark}.}  Yet the findings in this paper suggest otherwise.  Achieving equality of opportunity may remain elusive even in the absence of psychological prejudice, historical wrongs, and differences in ability or education among racial groups.\footnote{Notwithstanding little change in recent support for redistribution despite rises in inequality \parencite{ashok2015}, greater understanding of the fairness (or lack thereof) of the economic system might influence some people's preferences for redistribution \parencite{alesina2018}.}  Advantages from being within a larger or more strongly connected social network may persist despite one's talent.  All else equal, outcomes may remain unequal. \par

This paper makes four main contributions.  First, the paper presents a theoretical approach for uncovering discriminatory outcomes independent from discriminatory motives (that is, independent from both prevailing models of discrimination in economics: taste-based and statistical)---unlike in previous work, these discriminatory outcomes arise even under an initial state of equality between majority and minority groups.  In other words, not only does this paper develop a standard labor-market model that reveals the limitations of past economic models of discrimination, but also this work offers a direct rigorous account of \textit{social network discrimination} using mainstream economic theory.\footnote{In this paper, one way social network discrimination manifests is as minorities receiving fewer job referrals, all else equal, simply because their social group is smaller.  Though capturing a different concept, social network discrimination may relate to ``institutional discrimination'' in certain contexts, which \textcite{small2020} defines as differential treatment by race that is either perpetrated by organizations or codified into law.  Yet the social network discrimination introduced in this paper does not rely on organizations beginning as racially homophilous, unlike the example of institutional discrimination their paper presents.  Their paper also provides a lengthy discussion on the divergence between economic and sociological approaches to discrimination.}  Second, the model in the paper predicts that in the presence of race-blind hiring policies, minorities will be inherently disadvantaged, all else equal.  This prediction applies to cases in which social structure and human capital are initially equal between majority and minority workers.  According to Proposition 1B, the prediction of minority disadvantage can be mitigated or eliminated if the minority group begins with sufficient compensatory advantages over the majority group---for example, sufficiently \textit{more} social ties or a sufficiently more tightly-knit social network.  Third, this paper performs equilibrium analysis of employment and wage differences caused by homophily along majority/minority status, which is a contribution to both the economic and sociological literatures on labor markets.  (I review the literature and make more specific claims about this paper's contribution to the existing literature later in this Introduction.)  In doing so, the paper isolates a potential underlying mechanism for inequality, adding to our understanding of racial disparities that have been widely studied across the social sciences.  Fourth, the paper calibrates the model using a nationally representative sample from the National Longitudinal Study of Adolescent to Adult Health.  In this calibration exercise, white respondents represent the majority group, and black respondents represent the minority group.  Through calibration, I estimate that the lower bound welfare gap---that is, the minimum difference in expected wages---caused by social network discrimination is up to 3.2 percent, with black respondents disadvantaged compared to white ones.  To the best knowledge of the author, there has not been any prior quantification of the magnitude of impact social network discrimination has in isolation on racial disparities.

\textbf{Contributions to Literature.}
The model in this paper extends the one in \textcite{montgomery1991} to incorporate two-dimensional heterogeneity: while the original model only groups workers by ability, this paper also groups them by majority/minority status.  Doing so yields findings that go far beyond a mere application of the base model.  The original model does not focus on outcomes when homophily (the well-documented tendency for people to associate more with others similar to themselves) exists along characteristics \textit{uncorrelated} with ability---namely, on what effects emerge when social ties are formed along dimensions orthogonal to productivity.  The original model does not incorporate demographic considerations. \par

Filling this gap is important.  Within sociology, research has explored homophily in various contexts, including its causes \parencite{wimmer2010, leszczensky2019} as well as how it influences friendships \parencite{blau1977, syed2012}, interethnic marriages \parencite{skvoretz2013}, and social inequality \parencite{dimaggio2012}, among other areas.  \textcite{rubineau2013} shows how referral behavior can segregate jobs beyond the influence of homophily.  Within economics, there is an extant literature on homophily \parencite[see, for example,][]{golub2012}, and on the impact of referrals on inequality,\footnote{For some of the evidence, see \textcite{arrow2004, ioannides2004, bayer2008, hellerstein2011, renneboog2011, burks2015, pallais2016, chandrasekhar2020, miller2021}.  In a different setting than referrals, \textcite{dasaratha2020} finds disproportionate benefits from peer effects due to homophily.} yet many inequality findings have relied on the existence of some degree of prior period inequality beyond majority/minority group size---for example, that if a demographic group has higher past employment, then that would yield an advantage in securing future jobs.  This paper adds to insights from both fields, demonstrating that referral advantages may still unequally accrue over time even under initial equality between majority/minority groups, due to homophily.  In particular, this paper adds a theoretical foundation for why homophily may contribute to inequality in referral markets,\footnote{In contrast, \textcite{zeltzer2020} presents empirical evidence to suggest that gender homophily is a significant factor in explaining the gender-wage gap among medical professionals.} as well as predicts under what conditions such disparities will \textit{not} surface.  \par

There has been increasing focus on uncovering mechanisms behind racial disparities in labor market outcomes \parencite{bayer2018, chetty2020} and wage inequalities \parencite{card1994changing, lemieux2006increasing}.  Evidence for racial differences in networking outcomes exists \parencite{korenman1996, lalanne2011, mengel2015, lindenlaub2016}.  \textcite{jackson2009} discusses how homophily leads to segregation of groups, which leads to different equilibrium investment decisions in areas like education.  Yet this paper introduces a more direct mechanism for inequality from homophily: that even given \textit{equal} investments in human capital, homophily may still directly foster disparities through hiring dynamics.\footnote{Early versions of these findings can be found in the author's undergraduate honors thesis (see \textcite{okafor2007}).  \textcite{buhai2020} explores how inequality can result from \textit{different} choices in skill specialization, that is, different investments in human capital.}  \textcite{bolte2020} finds that inequality due to homophily arises as a function of historical employment, whereas this paper does not rely on historical differences in labor market conditions.  Similarly, unlike \textcite{calvo2004}, which provides a network-based mechanism for inequality deriving from differential drop-out rates, this paper uncovers how homophily can lead to inequality through the channel of employment opportunities themselves.  Thus, this paper introduces a \textit{direct} mechanism through which homophily can generate inequality that does not rely on historical differences between social groups, a contribution not previously found in the literature to the best knowledge of the author.  In addition, this paper's findings add to ones from sociology that link homophily to social inequality \parencite[see, for example,][]{dimaggio2012}; in this paper, in contrast, homophily's impact on inequality does not operate through a mechanism that exacerbates individual-level differences (recall our model assumes groups have \textit{equal} ability and initial employment).  Notably, this paper's findings are also distinct from past economics and sociology research on the influence of more traditional discrimination in hiring \parencite[see, for example,][]{bertrand2004, pager2009}.  Unlike those papers, this one uncovers disparities even in contexts in which discriminatory motives and implicit biases are not only absent but impossible, as the model in this paper does not allow firms to distinguish who is a majority and who is a minority worker.  Significantly, if the biases suggested by such prior research are incorporated---biases that tend to favor the majority group---the size of the disparities predicted in this paper's model would be even larger.\par

This paper proceeds as follows: Section \ref{sec:modelassumptions} introduces a formal setup of the model.  Section \ref{sec:modelresults} presents the model's key findings for majority and minority workers.  Section \ref{sec:modeldiscussion} provides discussion, including a description of (1) why the particular parameterization of homophily is used and is consistent with the contact hypothesis from psychology; (2) why other parameterizations are not used; (3) model considerations in light of historical context; and (4) model considerations in the university context. Section \ref{sec:modelcalibration} performs a calibration of the model under simplifying assumptions to estimate the lower bound magnitude of social network discrimination in real-world settings.  Section \ref{sec:modelconclusion} concludes by introducing some of the key theoretical, practical, and legal implications of the findings.

 \par

 \par

\section{Model} \label{sec:modelassumptions}

Here, I extend the \textcite{montgomery1991} two-period model to incorporate two-dimensional heterogeneity: while the original model only groups workers by ability, this one also groups them by majority/minority status.\footnote{As \textcite{montgomery1991} states, most of the model's assumptions are ``standard in labor-market models of adverse selection,'' especially that of \textcite{greenwald1986}.} \par 

\textit{Workers:} I consider a labor market with two time periods ($t=1$ and $t=2$) and many workers, with an equal measure in each period.\footnote{Similar to \textcite{montgomery1991}, I simplify the analysis by examining the model as the number of workers approaches infinity.}  Each worker works one period, and is one of two types: majority or minority.  Each worker's type is predetermined and assigned before the period in which he or she enters the market.  The fraction of majority workers is $\delta  >\frac{1}{2}$, while $1 -\delta  <\frac{1}{2}$ are minority.  Similar to \textcite{montgomery1991}, I assume that $\frac{1}{2}$ of the workers within each type are high-ability, while $\frac{1}{2}$ are low-ability.   High-ability workers produce one unit of output, while low-ability workers produce zero units.  Workers are observationally equivalent: firms neither know what ability workers possess (before production), nor whether workers are of the majority or minority type (at any time).\footnote{If the majority/minority type is based on race, then this environment would correspond with a ``race-blind'' setting.  Later we see that period-1 workers' actions are nonstrategic; hence, no assumption needs to be made on their knowledge of their own or of period-2 workers' types.}\par

\textit{Firms:} Firms are free to enter the market in either period.  At most, each firm may employ one worker.  A firm's profit in each period is equal to the productivity of its worker minus the wage paid.\footnote{Product price is exogenously determined and normalized to unity.}  Each firm must set wages before it learns the productivity of its worker.  There are no output-contingent contracts.\footnote{This assumption captures a significant rationale for screening of job applicants and the use of referrals: the inability to fully tie compensation to productivity.  See \textcite{montgomery1991, greenwald1986} for further discussion of this assumption.} In period-1, firms hire workers through the market.  In period-2, firms hire workers either through referrals or through the market.  \par

\textit{Timing:} Each firm hires a period-1 worker through the market; after production, they learn the worker's ability.  Since period-1 workers are observationally equivalent (and cannot be referred for jobs since there is no previous time period), each firm that hires through the market receives a high-ability worker with probability $\frac{1}{2}$. \par

After each firm learns the ability of its current worker, they may set a referral offer to be communicated to their worker's social tie.  If a firm chooses not to set a referral offer, it can go to an anonymous pool (the period-2 market).  This period-2 market is comprised of either workers who received no referrals in the prior period, or who did not accept any of the referrals they did receive in the period-1 market.  Firms who do hire in the period-2 market may hire a worker picked at random in this period-2 market. \par

If a firm sets a referral offer, whether the referral offer is relayed is conditional on the firm’s worker holding a social tie.  If he or she does hold one, then the firm will only attract the worker's social tie if the referral offer exceeds both their outside option (the period-2 market wage) and all other referral offers received by the worker's social tie.  A firm that does not wish to hire through referral will not set a referral offer (or might simply set a referral offer below the period-2 market wage, which automatically has no chance of acceptance).  Period-2 workers then compare all of the referral offers they receive, choosing the highest. \par

All period-2 workers who do not receive any referral offers must find employment through the general market. \par

In summary, the timing of the game is as follows.
\begin{enumerate}
    \item Each firm hires period-1 workers through the period-1 market at a wage of $w_{M1}$.
    \item Period-1 production occurs, after which all firms learn the productivity of their worker.
    \item Social ties are determined (the process of social tie formation is described below).
    \item If a firm wishes to hire through referral, it sets a referral offer.  I denote firm $i$'s referral offer by $w_{Ri}$.  (Conditional on having a social tie, each period-1 worker then relays his or her firm's wage offer ($w_{Ri}$) to their period-2 acquaintance.)
    \item Each period-2 worker compares all wage offers received.  They either accept one or wait to find employment through the general market.
    \item Any period-2 worker with no offers (or who refuses all offers) goes on the market.  Wages in this market are denoted $w_{M2}$.
    \item Period 2 production occurs.
\end{enumerate}

\textit{Structure of Social Network:} As the focus of the model is referrals, now I describe how the social network is drawn through which referrals occur.  As described, there are four categories of workers: high-ability majority, high-ability minority, low-ability majority, and low-ability minority.  

To draw the social network, the following steps can be taken for each period-1 worker: first, determine if the period-1 worker has a social tie, which occurs with a probability equal to the \textit{network density} for their respective majority or minority type (denoted by $\tau _{maj}$ and $\tau _{min}$, respectively).\footnote{If the period-1 worker is a majority worker, he or she possesses a social tie with probability $\tau _{maj} \in (0 ,1)$; a minority worker possesses a social tie with probability $\tau _{min} \in (0 ,1)$.}  Second, conditional on the period-1 worker having a social tie (determined based on step 1 immediately above), their social tie can be randomly assigned to a period-2 worker who is both of:\footnote{The ``urn-ball'' model is standard in probability theory and can also be conceptualized to understand how the social network is drawn (for more background on the ``urn-ball'' model, see \textcite{shimer2007}).  Specifically, we can represent each period-2 worker as an urn, and each social tie that a period-1 worker possesses as a ball.  The assignment of social ties is equivalent to a scenario where the balls are randomly dropped into the urns.  A period-1 worker's sole social tie, if they have one, is dropped into an ``urn'' (period-2 worker) which is: (1) of the same ability with probability $\alpha  \in (\frac{1}{2}, 1)$; and (2) of the same majority/minority type with a probability determined by the period-1 worker's in-group bias.  Period-2 workers can have social ties with zero, one, or more than one period-1 worker.} (1) the same high or low ability as the period-1 worker with a probability equal to the \textit{ability in-group bias} (denoted by $\alpha \in (\frac{1}{2},1)$)\footnote{The matched period-2 worker is hence of a different ability with probability $1 - \alpha  \in (0, \frac{1}{2})$.}; and (2) the same majority or minority type as the period-1 worker with a probability calculated directly from their \textit{type in-group bias} (denoted by $\psi _{maj}$ and $\psi _{min}$ for the majority and minority type, respectively).  The mathematical relationship between the type in-group bias parameter and the likelihood of having a social tie with the same majority/minority type is:
         \begin{align*}
                \Pr\{\text{social tie with same majority/minority type}\} =\frac{w \cdot \psi}{[w \cdot \psi] + [(1-w)(1-\psi)]},
            \end{align*}
        where $w$ is the share of the labor force for the period-1 worker type (either $\delta$ for the majority type or $1- \delta$ for the minority type) and $\psi$ is the type in-group bias of the period-1 worker's type (either $\psi _{maj}\in (\frac{1}{2}, 1)$ or $\psi _{min}\in (\frac{1}{2}, 1)$).\footnote{A simple example for why this expression is used can be found in the next Subsection (see Equation \ref{eq:biasmaj} and Equation \ref{eq:biasmin}).  In addition, Equation \ref{eq:biascombined} and the Discussion Section provide more explanation of this mathematical expression.}

The network structure is thus characterized by four parameters: network density ($\tau _{maj}$ and $\tau _{min}$ for majority and minority workers, respectively), ability in-group bias (denoted by $\alpha$), type in-group bias (denoted by $\psi _{maj}$ and $\psi _{min}$ for majority and minority workers, respectively), and the majority share of workers ($\delta$). A summary of the model parameters can be found in Table \ref{tab:modelparameters}.\par

\textit{Note on Type In-Group Bias:} Type in-group bias ($\psi_{maj}$ and $\psi_{min}$) can be conceptualized as capturing the fact that there are shared attributes that simply make it easier for some workers to form social ties with each other than with others.  For example, for a given chance encounter, one is more likely to form a social tie with another worker who shares a more similar background, because there are simply more elements in common to establish the foundation of a relationship---for example, consider the natural bonding that occurs when two people meet who grew up in the same neighborhood or who experienced the same cultural traditions.  Hence, type in-group bias does not represent favoritism toward a demographic group: in the model, any given worker views members of one's own majority/minority group and members of the other group equivalently, conditional on having a social tie.  Similarly, conditional on \textit{not} having a social tie, any given worker views members of the same majority/minority group and members of the other group with the same level of (dis)interest.  Hence, although type in-group bias deeply impacts network formation, it is fully distinct from (racial or group) animus, in-group favoritism, or traditional conceptions of taste-based preferences in economics.  

\begin{ThreePartTable}
\begin{TableNotes}[flushleft]
\footnotesize
\item[*] I assume neither full bias nor no bias.  Furthermore, $\psi \in [0, \frac{1}{2})$ represents heterophily, a rare social network phenomenon outside the scope of this paper.
\end{TableNotes}

\begin{spacing}{1}
\begin{longtable}{cccc} 
\caption{Model Parameters}\\
\toprule
Parameter & Name & Description & Range \\
\midrule
\endfirsthead

\caption[]{Model Parameters (Continued)}\\
\toprule 
Parameter & Name & Description & Range\\
\midrule
\endhead

\bottomrule
\endfoot

\bottomrule
\insertTableNotes
\endlastfoot

$\delta$ & Majority share & \makecell{Share of the total labor force \\ comprised of majority workers} & $\delta \in (\frac{1}{2}, 1)$ \\[1em] \\

$1 - \delta$ & Minority share & \makecell{Share of the total labor force \\ comprised of minority workers} & $1 - \delta \in (0, \frac{1}{2})$ \\[1em] \\

$\alpha$ & \makecell{Ability \\ in-group bias} & \makecell{Probability a worker's social tie is \\ with another worker of equal ability} & $\alpha \in (\frac{1}{2}, 1)$ \\[1em] \\

$\tau_{maj}$ & \makecell{Network density \\ (majority group)} & \makecell{Probability a majority worker \\ has a social tie} & $\tau_{maj} \in (0, 1)$ \\[1em] \\

$\tau_{min}$ & \makecell{Network density \\ (minority group)} & \makecell{Probability a minority worker \\ has a social tie} & $\tau_{min} \in (0, 1)$ \\[1em] \\

$\psi_{maj}$ & \makecell{Type \\ in-group bias \\ (majority group)} & \makecell{Bias a majority worker exhibits \\ toward workers of the same group \\ in favor of forming social ties. Value \\ of 1/2 means no bias (probability \\ of social tie with another majority \\ worker = $\delta$). Value of 1 means \\ full bias (probability of social tie \\ with another majority worker = 1).*} & $\psi_{maj} \in (\frac{1}{2}, 1)$ \\[1em] \\

$\psi_{min}$ & \makecell{Type \\ in-group bias \\ (minority group)} & \makecell{Bias a minority worker exhibits \\ toward workers of the same group \\ in favor of forming social ties. Value \\ of 1/2 means no bias (probability \\ of social tie with another minority \\ worker = $1 - \delta$). Value of 1 means \\ full bias (probability of social tie \\ with another minority worker = 1).*} & $\psi_{min} \in (\frac{1}{2}, 1)$

\label{tab:modelparameters}

\end{longtable}
\end{spacing}
\end{ThreePartTable}

\subsection{Implied Model of Referrals}
\noindent A simplified implied model of referrals from the formal exposition of the social network structure above is as follows.  First, if asked to provide a referral by the employer, the employed worker in period 1 provides the referral offer to their social tie, if they have a tie. Second, the period-1 worker has a social tie with probability $\tau_{maj}$ (if period-1 worker is majority) or $\tau_{min}$ (if period-1 worker is minority). Third, the worker's social tie is determined from their direct contacts. Fourth, the pool of direct contacts for a period-1 high-ability \textit{majority} worker is determined as follows:
    \begin{enumerate}[label=\alph*.]
        \item First, sample potential contacts at random from the population.
        \item For each potential contact, if they are also from the majority, keep them with probability $\psi_{maj}$.  If they are in the minority, keep them with probability ($1-\psi_{maj}$).
        \item The expected share of majority contacts is therefore:
            \begin{align} \label{eq:biasmaj}
                \phi_{maj}=\frac{\delta \cdot \psi_{maj}}{[\delta \cdot \psi_{maj}] + [(1-\delta)(1-\psi_{maj})]}
            \end{align}
        \item Finally, for each contact, if they are of high ability, keep them with probability $\alpha$.  If not, keep them with probability $1-\alpha$.\footnote{Since high-ability and low-ability workers each occupy $\frac{1}{2}$ of the population, the expected share of same-ability contacts simply equals $\alpha$.}
        \item The social tie is determined at random from their pool of contacts.
    \end{enumerate}
Fifth, the pool of direct contacts for a period-1 high-ability \textit{minority} worker follows the same steps (a)-(e), except for (b) and (c), which are modified accordingly:
    \begin{enumerate}[label=\alph*.]
        \setcounter{enumi}{1}
        \item For each potential contact, if they are also from the minority, keep them with probability $\psi_{min}$.  If they are in the majority, keep them with probability ($1-\psi_{min}$).
        \item The expected share of minority contacts is therefore:
            \begin{align}\label{eq:biasmin}
                \phi_{min}=\frac{(1-\delta) \cdot \psi_{min}}{[(1-\delta) \cdot \psi_{min}] + [\delta \cdot (1-\psi_{min})]}
            \end{align}
    \end{enumerate}
Sixth, the probability that a high-ability majority worker has a social tie with another high-ability majority worker is therefore $\tau_{maj}\phi_{maj}\alpha$.  Similarly, the probability that a high-ability minority worker has a social tie with another high-ability minority worker is $\tau_{min}\phi_{min}\alpha.$

\section{Results} \label{sec:modelresults}

The first subsection below characterizes the equilibrium.  The second subsection presents properties shared with \textcite{montgomery1991}, while the third subsection presents new propositions that I have added to the model by incorporating two-dimensional heterogeneity (that is, categorization of workers by majority and minority type in addition to (instead of solely by) ability level).  These new propositions relate to discrimination and labor market disparities.

\subsection{Characterization of Equilibrium}
The equilibria of the game are (weak) perfect Bayesian.\footnote{The exposition of the equilibrium characterization in this subsection is modified from \textcite{bolte2020}.}

Backward induction helps illustrate the equilibrium structure.  Period-2 workers will accept any wage that is at least $w_{M2}$, since that is also their option if they forego the referral market.  The only possibility for workers to mix strategies at $w_{M2}$ is if both workers and firms equally value the available strategies being mixed.\footnote{Otherwise, firms could deviate and offer a higher wage if workers mixed strategies, which would mean there would not exist an equilibrium in this scenario.}  If that is the case, the mixing does not matter since there is the same expected value for both workers and firms (which is 0), regardless of the action.  Thus, equilibrium behavior for hiring in the general market is such that firms hire period-2 workers from the general market at a wage of $w_{M2}$ if the expected productivity of workers in that market exceeds $w_{M2}$, do not hire from the general market if expected productivity is below $w_{M2}$, and both sides can mix arbitrarily if the expected productivity of workers in the market is exactly $w_{M2}$ (which is also the workers' option if they forego the referral market).\par

Given the strategies of others---along with the designated continua of workers and firms---firms have a weakly positive expected value from waiting and hiring from the general market.  A firm prefers hiring a period-2 worker who is referred if and only if that worker's productivity minus the referral wage is higher than the expected value from the period-2 general market.\footnote{Since worker productivity is either 0 or 1 in this model, Lemma 2 shows this condition is satisfied if the worker is a high-ability worker (that is, has a productivity of 1).}  The expected productivity of the period-2 worker is a function of the productivity of their social tie, the period-1 worker making the referral.\footnote{Specifically, the expected productivity of a period-2 social tie from a high-ability period-1 worker is $\alpha$, whereas the expected productivity of a period-2 social tie from a low-ability period-1 worker is $1-\alpha$.}  In an alternative setting without competition for the referred period-2 worker, the worker's wage would be $w_{M2}$ instead. \par

The equilibrium can thus be characterized by a threshold such that each firm seeks to hire a referred period-2 worker when the period-1 worker has a productivity above the threshold, and does not hire via referral below the threshold.  That threshold is linked to the firm's value from instead waiting to hire from the general period-2 market.  Let $i$ denote a generic period-1 worker, $a_i$ denote the productivity of the period-1 worker, $q$ denote a generic period-2 worker, and $v_q$ denote the productivity of the period-2 worker.  As modified from \textcite{bolte2020}, the threshold is characterized by a function that calculates the expected value of period-2 workers in the general market conditional on: (1) firms hiring via referral from period-1 workers with productivity strictly above $\widetilde{a}$, (2) not hiring via referral from period-1 workers with productivity strictly below $\widetilde{a}$, and (3) hiring via referral from period-1 workers with productivity exactly equal to $\widetilde{a}$ with probability $r$:%
\begin{align} \label{eq:equilibrium}
\begin{split}
    E_{\widetilde{a},r}[v_q|q&\in \text{market}]\}:= \\ &\frac{\begin{aligned}
        P(0) &\times && E[v_q]\text{ }+\\
        (1-P(0)) &\times && (\Pr(a_i < \widetilde{a})E[v_q | \text{ } a_i <\widetilde{a} \text{ }\cap \text{ } i \in <q,i>] \text{ } + \\
        & &&\Pr (a_i=\widetilde{a})(1-r)E[v_q | \text{ } a_i =\widetilde{a} \text{ } \cap \text{ } i \in <q,i>])\end{aligned}}{P(0) + (1-P(0))(\Pr(a_i < \widetilde{a}) + \Pr(a_i = \widetilde{a})(1-r))}
\end{split}
\end{align}
            
\noindent where $<\bullet\text{ }, \text{ } \bullet>$ denotes pairs of workers for whom a social tie is present, $P$ denotes the distribution of referral job offers a period-2 worker gets, with $P(k)$ being the likelihood that a worker gets precisely $k$ referral job offers.  Therefore, each equilibrium is equivalent to using a threshold $\widetilde{a}$ (along with a mixing parameter $r$), which can be derived from a fixed point of Equation \ref{eq:equilibrium}.  The following lemmas further characterize the equilibrium.

\subsection{Basic Equilibrium Properties}
Basic equilibrium properties are shared with \textcite{montgomery1991} and can be expressed via the following three lemmas, which establish that (1) referral wage offers are dispersed within an interval between the period-2 market wage and a maximum referral wage offer; (2) a firm will only hire through referral if it employs a high-ability worker in period 1; and (3) the period-1 market wage is greater than the expected period-1 productivity.  One important implication from point (2) is that it means the expected productivity of the referral market will be higher than that of the general period-2 market, and---relating to point (1)---it can be easily shown that the average referral wage will be higher than the period-2 market wage as a result.  More discussion on each of these points is included below.  Omitted proofs are found in Appendix \ref{sec:appendixa}.

\begin{lemma}
In equilibrium, referral wage offers lie within the interval between $w_{M2}$ and a maximum referral wage offer $\overline{w}_{R}$; hence, \emph{$w_{R} \in [w_{M2}$, $\overline{w}_{R}$]}.  The density of the referral wage offer distribution is positive across this entire range.
\end{lemma}

As \textcite{montgomery1991} also finds, the market wage ($w_{M2}$) must coincide with the bottom of the referral wage distribution, as any referral offer below the market wage will necessarily be rejected by workers in favor of their outside option (going to the general market).  At the maximum referral wage offer (denoted $\overline{w}_{R}$ and derived in Appendix Equation \ref{eq:maxreferral}), the probability a worker accepts the referral offer is 1 (as there are no better wage offers in either the referral market or general market).  Hence, firms will not offer a referral wage above this amount as it will necessarily reduce profits.  As \textcite{montgomery1991} also finds, if there were a gap in the wage distribution, say between some $w_1$ and $w_2$, then a firm offering the higher wage could reduce its offer by $\epsilon$ without reducing the probability its offer is chosen, which would increase its profits.

\begin{lemma}
A firm will attempt to hire through referral if and only if it employs a high-ability worker in period 1.
\end{lemma}
This result follows from the ability in-group bias ($\alpha$).  As \textcite{montgomery1991} also finds, hiring through the market yields zero expected profit (due to the free entry of firms and the symmetric lack of information about worker ability).  For firms employing high-ability workers in period 1, an accepted referral offer yields constant positive profit over the range of the referral offer distribution $[w_{M2} ,\overline{w}_{R}]$.  Higher wage offers yield a higher probability of attracting a period-2 worker.  Firms employing low-ability workers in period 1 will not hire through the referral market, since ability in-group bias (previously denoted by $\alpha \in (\frac{1}{2},1)$) means that the social tie of a low-ability worker is more likely than not to be with another low-ability worker.
\par 
As a result of this lemma, a disproportionately high number of low-ability workers find employment through the general market.  This drives the market wage below the average productivity of the entire population.  However, as \textcite{montgomery1991} states, adverse selection does not eliminate the market.  Since some high-ability workers are not ``well-connected,'' they fail to receive referral wage offers, which leads them to find employment in the general market.  Thus, the market wage remains above zero. \par

\begin{lemma}
The period-1 market wage is greater than the expected period-1 productivity.
\end{lemma} 
As \textcite{montgomery1991} also finds, if a firm obtains a high-ability worker in period 1, it expects positive period-2 profits.  This fact drives the period-1 market wage higher than the productivity of the population.  This wage can be viewed as comprising the average productivity of the worker plus an ``option value'' of a period-2 referral.  This option will be exercised if the period-1 worker reveals themselves to be high-ability (which occurs after period-1 production concludes if the worker is in fact high-ability).

\subsection{Propositions on Social Network Discrimination}
\label{sec:propdisc}

New propositions reflecting social network discrimination and inequality are detailed below.\footnote{An interactive plot that shows how social network parameters influence the distribution of referral offers received between majority and minority groups can be found at \href{https://justiceinsights.shinyapps.io/social_network_discrimination/}{https://justiceinsights.shinyapps.io/social\_network\_discrimination/}}  The propositions establish that minority workers receive a disproportionately low fraction of job offers through referral and a lower expected wage, all else equal.  Recall that the market wage lies below the referral offer distribution.  Hence, all these effects taken together yield a welfare gap between minority and majority workers in period 2 that did not exist in period 1, disadvantaging minority workers.  Omitted proofs are in the Appendix. \par

\setcounter{proposition}{1}
\begin{subprop}
In an environment with equal magnitude of majority/minority network parameters ($\tau_{maj}=\tau_{min}$ and $\psi_{maj}=\psi_{min}$), the probability that, among referral offers, a minority worker is referred is lower than their share of the labor force.  Conversely, the probability that, among referral offers, a majority worker is referred is larger than their share of the labor force.
\end{subprop}
\begin{subprop}
The inequality in the distribution of referral job offers can be eliminated by minority workers having a sufficiently higher type in-group bias ($\psi_{min}$).
\end{subprop} 

\paragraph{Simplified Numerical Examples:} This paper includes several separate sources for insight into the intuition behind Proposition 1A:

\begin{enumerate}
    \item A ``real-world'' scenario from Figure \ref{fig:intuition} of the Introduction involves trusted employees interacting with prospective job applicants at a professional event.\footnote{A video walking through this scenario can be found online at \textcite{okaforyoutube}.}
    \item A more mathematical simplified example is included below:
\end{enumerate}

First, I explore the intuition of Proposition 1A by using a simple numerical example that relaxes some of the assumptions of the model to illustrate how type in-group bias operates.\footnote{This example is illustrative and does not incorporate ability heterogeneity (that is, it assumes all workers are high-ability), so is simplified relative to the main model.  Furthermore, the example includes a finite number of workers whereas the model approaches a continuum.}  Suppose there are two majority workers in both periods and one minority worker in both periods---namely, $\delta =2/3$.  Also suppose there is type in-group bias: each worker has a bias in favor of forming social ties with workers of the same majority/minority type.  To illustrate this bias, let us say that for encounters between majority period-1 and majority period-2 workers, there is a 2/3 chance of forming a tie, whereas encounters between a majority period-1 and minority period-2 worker have a 1/3 chance of forming a tie---that is, $\psi = 2/3$.  Suppose all three period-1 workers encounter all three period-2 workers.  The expected number of ties period-1 majority workers form with their own type is thus 4/3 (while the expected number of ties with minority workers is 1/3); this means the fraction of same-type social ties for majority workers is 4/5.  Let $\phi_{maj}=0.8$ denote this fraction of same-type social ties (note that $\phi$ is not a parameter of the model and can be calculated directly from $\psi$ via Equation \ref{eq:biascombined} below).  Similarly, it is straightforward to calculate that the fraction of same-type social ties for the period-1 minority worker is $\phi_{min}=0.5$. \par

Now let us apply these bias dynamics to a case that more closely conforms with the assumptions of the model---in which period-1 workers each have exactly one social tie with probability 1 (that is, $\tau_{maj}=\tau_{min}=1$).  The fraction of referral job offers going to majority workers is simply a weighted sum:%
\begin{align*}
& \Pr\{\text{referral job offer goes to majority per-2 worker}\} \\
& = \Pr\{\text{per-1 worker is majority}\} \cdot \Pr\{\text{per-1 majority knows per-2 majority}\} \\
& \qquad + \Pr\{\text{per-1 worker is minority}\} \cdot \Pr\{\text{per-1 minority knows per-2 majority}\} \\
& = \delta \cdot \tau_{maj} \cdot \phi_{maj} + (1-\delta) \cdot \tau_{min} \cdot (1-\phi_{min}) \\
& = 2/3 \cdot 1 \cdot 0.8 + 1/3 \cdot 1 \cdot 0.5 \\
& = 0.7
\end{align*}%

Hence, only 0.3 of referral job offers go to minority workers, even though they occupy 1/3 of the labor force.  This simple example illustrates the distorting influence of having the same magnitude of bias operating on groups of different sizes.  The bias in favor of a same-type social tie for a majority worker extends toward a greater fraction of the population than does the same magnitude of bias for a minority worker.  Hence, equal magnitudes of bias \textit{unequally} impact the respective chances of knowing workers of the same type. \textbf{[End of Example]}%

I can generalize both the reasoning of the simplified numerical example above and the type-in group bias of the model as follows:%
\begin{align} \label{eq:biascombined}
 \Pr \{\text{period-1 worker knows own maj/min type}\} =\phi=\frac{w \cdot \psi}{[w \cdot \psi] + [(1-w)(1-\psi)]}
\end{align}

\noindent where $w$ is the share of the labor force for the worker type (either $\delta$ or $1-\delta$) and $\psi$ is the in-group bias of the worker type (either $\psi_{maj}$ or $\psi_{min}$).  $\psi = 0.5$ reflects no bias---that is, a proportional chance of a social tie being with another of the same type---and $\psi = 1$ reflects when all ties are with members of the same type.  $\phi$ represents either $\phi _{maj}$ or $\phi _{min}$, depending on whether the period-1 worker belongs to the majority or minority group, respectively.  The Discussion section further explores the relationship between $\psi$ and $\phi$, and its implications on the results.  \par

\begin{figure}
\captionsetup{justification=centering}
\caption{Magnitude of Minority Group Network Parameters \\ Required for Parity in Referral Job Offers}
\centering
\includegraphics{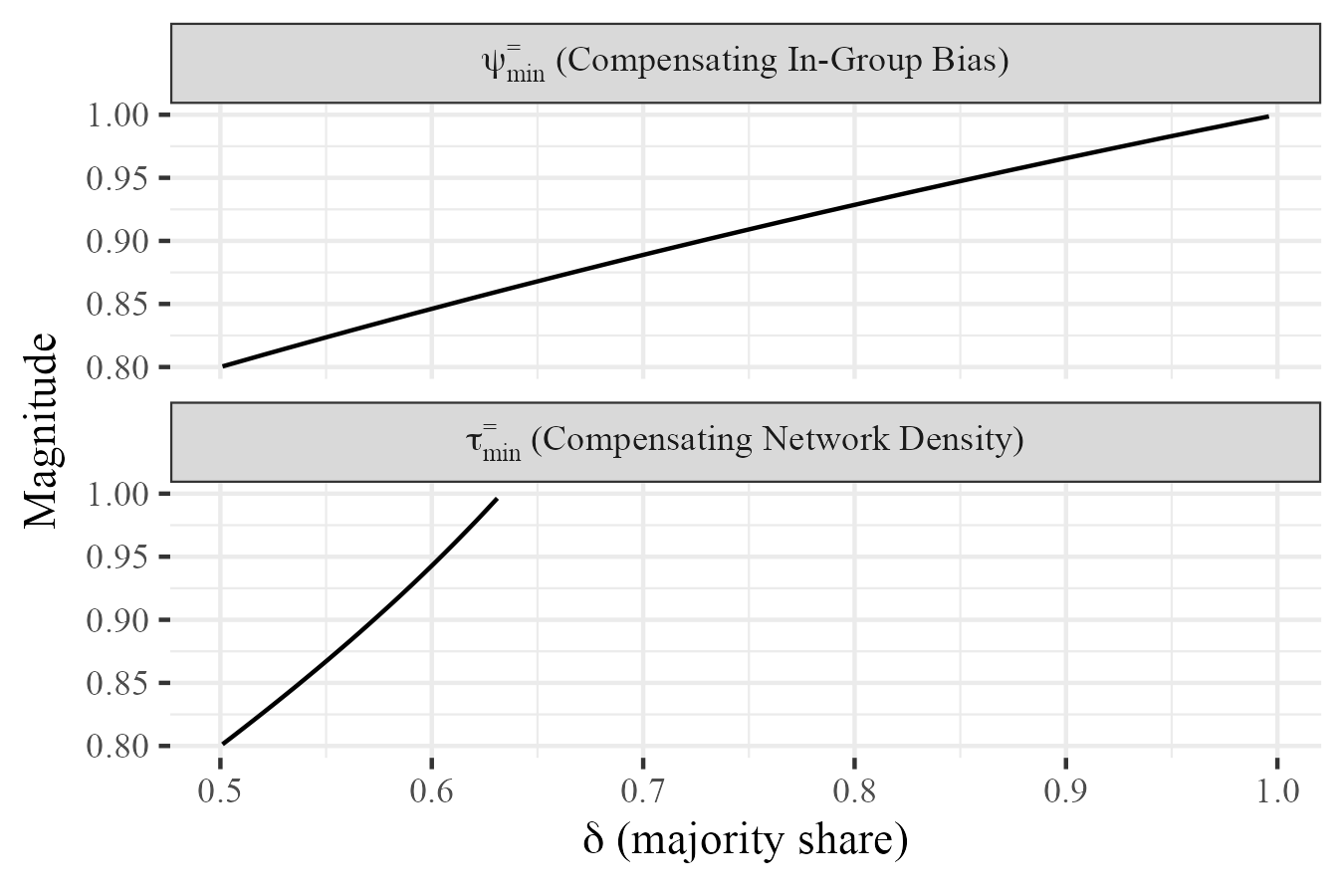} \\
\centering \footnotesize \textit{Note:} In both charts, all other relevant network parameters equal 0.8.
\label{fig:parity}
\end{figure}

In the Appendix, I prove that parity in the distribution of job offers between high-ability majority and minority workers is accomplished only when: 
\begin{align} \label{eq:parity}
        (1 -\delta )\left [\delta \tau _{maj} \phi _{maj} +(1 -\delta )\tau _{min}(1 -\phi _{min})\right ] =\delta \left [\delta \tau _{maj} (1 -\phi _{maj}) +(1 -\delta )\tau _{min}(\phi _{min})\right ]
\end{align}

\noindent where $\phi _{maj}$ and $\phi _{min}$ are calculated from Equation \ref{eq:biascombined} above.  From Equation \ref{eq:parity}, one can calculate the magnitude of the other parameters for parity.  I denote such parameters for minority workers as $\tau _{min}^{=}$ (compensating network density) and $\psi _{min}^{=}$ (compensating in-group bias).  Figure \ref{fig:parity} illustrates that a sufficiently high network density or type in-group bias can mitigate the disproportionality in the distribution of job offers through referral.  All else equal, minority workers can either have more social ties ($\tau _{min}^{=} > \tau _{maj}$) or a ``stronger-knit'' social network ($\psi _{min}^{=} > \psi _{maj}$). \par

The network density required to eliminate the inequality ($\tau _{min}^{=}$) increases in $\tau _{maj}$, $\psi _{maj}$, and $\delta$.  It decreases in $\psi _{min}$, which can readily be understood intuitively.  The greater the probability of majority workers having social ties (and/or the greater the degree of their homophily), the greater minority workers' compensating parameters ($\tau _{min}^{=}$ or $\psi _{min}^{=}$) must be to achieve a proportional amount of all job offers through referrals. In Figure \ref{fig:parity}, the plot of $\tau _{min}^{=}$ has no values when $\delta$ is greater than approximately 0.63.  This is because there is no attainable magnitude of network density that will yield parity in the distribution of job offers when $\delta$ surpasses this threshold.  \par

\begin{proposition}
In an environment with equal magnitude of majority/minority network parameters ($\tau_{maj}=\tau_{min}$ and $\psi_{maj}=\psi_{min}$), the period-2 market wage ($w_{M2}$) decreases as majority workers occupy a greater fraction of the labor force.  $w_{M2}$ also decreases in the ability in-group bias $\alpha$.
\end{proposition} 

Recall that workers who do not receive jobs through referral must find employment through the market.  Proposition 1 shows that minority workers, all else equal, receive a disproportionately low fraction of job offers through referral, and thus disproportionately find employment through the market.  Decreases in the market wage (\textbf{\textit{$w_{M2}$}}) thereby hurt the average welfare of minority workers relative to that of majority workers. \par

The Appendix includes the expression for $w_{M2}$.  Given $\alpha > \frac{1}{2}$,  $w_{M2}$ is always less than $\frac{1}{2}$, the average productivity of the population.  Analysis shows that $w_{M2}$ is decreasing in $\alpha$.  
For all $\psi _{maj} =\psi _{min}$ and $\tau _{maj} =\tau _{min}$, $w_{M2}$ also decreases in $\delta$. \par

\begin{proposition}
In an environment with equal magnitude of majority/minority network parameters ($\tau_{maj}=\tau_{min}$ and $\psi_{maj}=\psi_{min}$), the referral wage and welfare (namely, average expected wage) for minority workers are lower than for majority workers.
\end{proposition} 

Much of the intuition behind this finding follows from Proposition 1 (that majority workers receive a disproportionately high number of job offers through referral).  There are two margins to consider.  First, the extensive margin: majority workers disproportionately get hired through the referral market (which provides higher wages than the general market), driving up expected welfare for the majority group.  Second, the intensive margin: recall that workers accept the maximum referral wage offer received.  Hence, by majority workers receiving a higher number of referral offers, their expected maximum offer increases, thereby also driving up their relative welfare.\footnote{A simpler version of this model can be constructed in which wages are exogenously determined.  In such a case, it would be straightforward to prove that the equilibrium referral wage must be higher than the period-2 market wage, since the expected productivity of referred workers is higher than the expected productivity of workers who do not get hired through referral and therefore must find employment through the general period-2 market.  This difference in expected productivity follows directly from combining Lemma 2 with the fact that there is homophily in ability (that is, the ability in-group bias $\alpha \in (\frac{1}{2}, 1)$).  If one supposed that, to the contrary, in a model with exogenously determined wages, the referral wage and the period-2 market wage would be equal.  In such a case, workers would be indifferent between being hired in the referral market or in the period-2 general market.  This means a firm wanting to hire via referral could increase its profits by increasing its referral wage offer by $\varepsilon$ to ensure its referral wage offer would be accepted by any period-2 worker.  Thus, an exogenously-determined referral wage must be higher than the period-2 market wage.  Hence, in a model with exogenously-determined wages, Proposition 3 still holds: the prediction that minority workers have lower welfare compared to majority workers, all else equal, mechanically follows from Proposition 1A, in which minority workers receive disproportionately fewer job offers through referral---and thus receive disproportionately fewer job offers associated with the (higher) referral wage.} \par

Let  $E\prod _{H}(w_{R})$ denote the expected period-2 profit earned by a firm employing a high-ability worker and setting a referral wage.  Firms must earn the same expected profit on each referral wage offered to preserve equilibrium wage dispersion:%
\begin{align*}
        E\prod _{H}(w_{R}) =c\qquad  \forall w_{R} \in [w_{M2} ,\overline{w}_{R}]
\end{align*}

\noindent Given the expression for $c$ derived in the Appendix, firms with high-ability workers who have social ties receive positive expected profits as long as $\alpha > \frac{1}{2}$.  Analysis shows that $c$ is increasing in $\tau _{maj}$ and $\tau _{min}$. Furthermore, setting $E\prod _{H}(w_{R})$ equal to $c$ for all potential wage offers $w_{R}$ can allow one to determine the equilibrium referral-offer distribution $F(\bullet )$. \par

Unfortunately, doing so does not result in a closed-form solution for $F(w_{R})$.  As \textcite{montgomery1991} also finds, because there is a continuum of firms, the equilibrium referral-offer distribution $F(\bullet )$ can be interpreted as either: 1) each firm randomizes their referral offer over the entire distribution; or 2) a fraction $f(w_{R})$ of firms offers each referral wage with certainty.  From the second interpretation, I denote these referral wages with $w_{Rk}$.  One can then derive an expression for $w_{Rk}(\alpha ,\delta ,\tau _{maj} ,\tau _{min} ,\psi _{maj} ,\psi _{min} ,F(w_{Rk}))$ and calculate an average referral wage received by a majority worker (denoted $E(w_{R _{Hmaj}})$) vs. a minority worker (denoted $E(w_{R _{Hmin}})$), for any given $\delta$, $\alpha$, $\tau _{maj}$, $\tau _{min}$, $\psi _{maj}$, and $\psi _{min}$. \par

Analysis shows that when majority/minority network parameters are the same magnitude, if $\alpha > \frac{1}{2}$ and $\delta  >\frac{1}{2}$, $E(w_{R _{Hmaj}}) >E(w_{R _{Hmin}})$. In other words, the expected referral wage for high-ability majority workers is greater than for high-ability minority workers. \par

\begin{figure}
\captionsetup{justification=centering}
\caption{Estimated Welfare Gap of Minority Workers}
\centering
\includegraphics{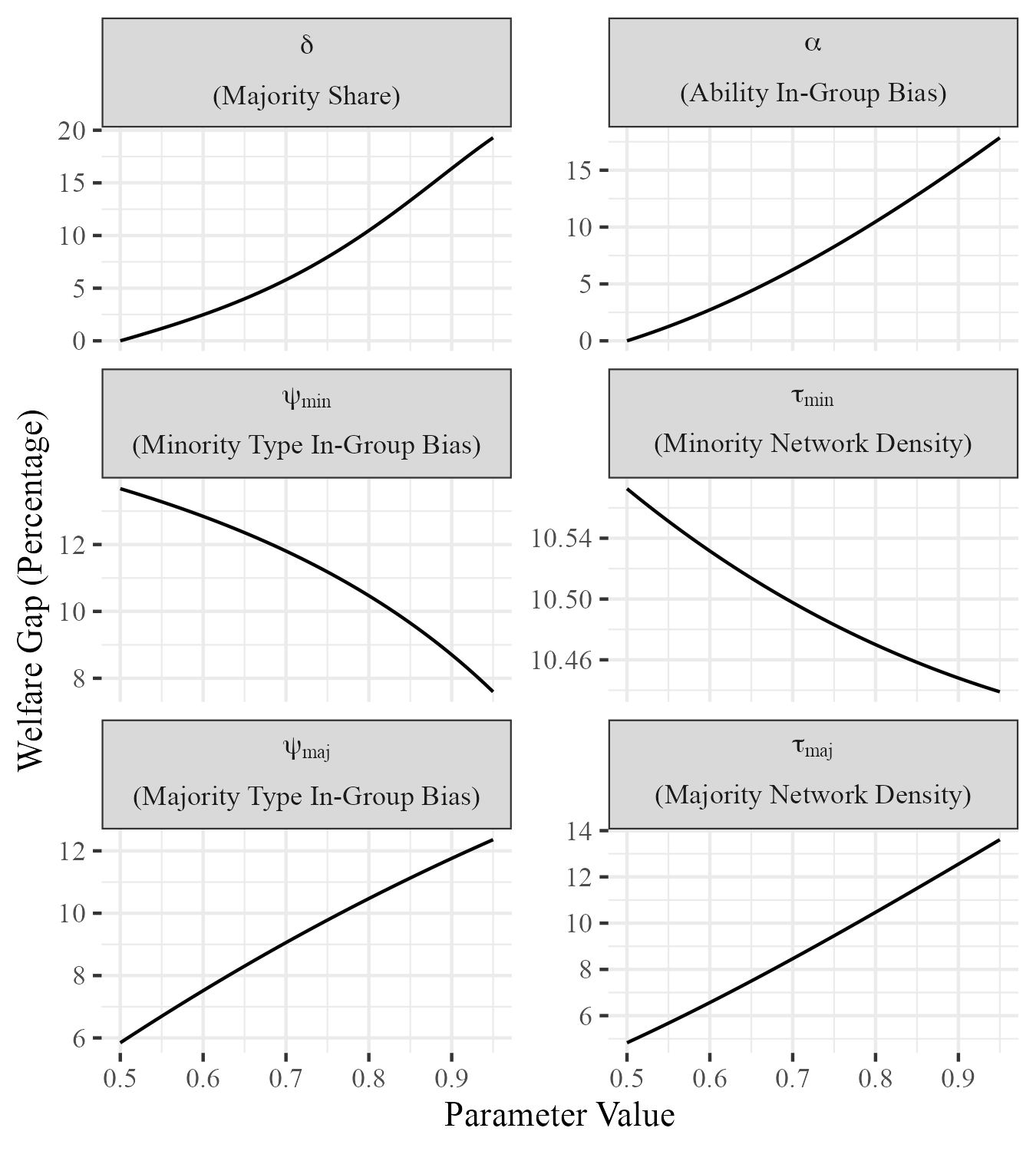}\\
\centering \footnotesize \textit{Note:} In each chart, the parameters not being varied all equal 0.8.
\label{fig:welfare}
\end{figure}

Welfare (that is, expected wage) is calculated by summing the market wage ($w_{M2}$) and expected referral wage ($E(w_{R _{Hmaj}})$ or $E(w_{R _{Hmin}})$), weighted by the likelihood of the worker's type gaining employment through the market or through referrals, respectively.  Figure \ref{fig:welfare} plots the estimated welfare gap between majority and minority workers as a function of various network parameters.\footnote{Estimation of wage gap normalizes to 1 the number of offers that referred high-ability minority workers receive and assumes a uniform distribution across the referral wage distribution ($w_{R} \sim U(w_{M2}$, $\overline{w}_{R}$)).}  The welfare gap increases in $\delta$, $\alpha$, $\psi_{maj}$, and $\tau _{maj}$.  It decreases in $\psi_{min}$ and $\tau _{min}$.  Of note, the welfare gap increases convexly as the majority group occupies a greater share of the labor force.

\section{Discussion} \label{sec:modeldiscussion}

If the majority/minority type in the model is based on race, then the setting would correspond with a ``race-blind'' or ``colorblind'' setting.  This is because workers are observationally equivalent, and so firms cannot incorporate race into hiring decisions.  Due to the historical context of race in the United States, past research has found bias \textit{in favor of} the majority group (see, for example, \textcite{bertrand2004}).  The presence of such bias would exacerbate the disparities already predicted by the model in this paper. The rest of this discussion section will first consider the choice behind the parameterization of homophily (that is, in-group bias).  Next, this section will consider why potential alternative parameterizations were not used.  Lastly, this section will consider the model's implications given the historical context of inequality.\par 

\subsection{Why This Homophily Parameterization is Used}

A key assumption of this paper's parameterization of homophily is that the probability of connection with a particular social group is increasing in that group's population share.  This assumption is reasonable due to at least two points: (1) the intuition introduced by the simplified numerical example following Proposition 1 in Section \ref{sec:propdisc}; and (2) empirical research on the contact hypothesis from psychology.  Regarding the first point: as the numerical example illustrates, since social connections form from interpersonal interactions, the more interpersonal interactions an individual has with members of a particular social group, the more social connections the individual is expected to form with that group.  Hence, the key assumption holds: the probability of connection with a particular social group would be increasing in that group's population share.  Regarding the second point: the key parameterization assumption is consistent with empirical research on the classic contact hypothesis from psychology.  The contact hypothesis suggests that prejudice and segregation are decreased through increased contact between in-group members and out-group members \parencite{allport1954}. Empirical research has found that as the quantity of contact with other social groups increases, cross-group relationships are fostered \parencite[see, for example,][]{hewstone2018, laurence2019}. Hence, the parameterization assumption that the probability of a connection with a particular social group is increasing in that group's population share holds.\par

Given the key assumption, this paper employs a parameterization of homophily that implies the relationship between $ \Pr \{$Period-1 worker knows own majority/minority type$\}$ and in-group bias is not linear.  Instead, the relationship depends on population share.  To understand why, consider an individual worker from each social group, as was done in the Simplified Numerical Example in Section \ref{sec:propdisc}.  For a majority worker, any given magnitude of bias in favor of same-type social ties should operate on a greater share of the population than it does for a minority worker since, by construction, a greater proportion of the population consists of majority workers.  Hence, one would expect the parameterization of homophily to have some multiplicative relationship with the share of the labor force, which a simple linear parameterization would not have.\footnote{See the Conclusion for several examples of common real-world settings in which one would expect these social network dynamics to take place.}   Furthermore, one would expect any given magnitude of homophily to have an amplified effect on the likelihood of having a same-type social tie for a given majority worker compared to a given minority worker. \par 

\begin{figure}
\caption{Social Tie Sensitivity to In-Group Bias}
\centering
\includegraphics{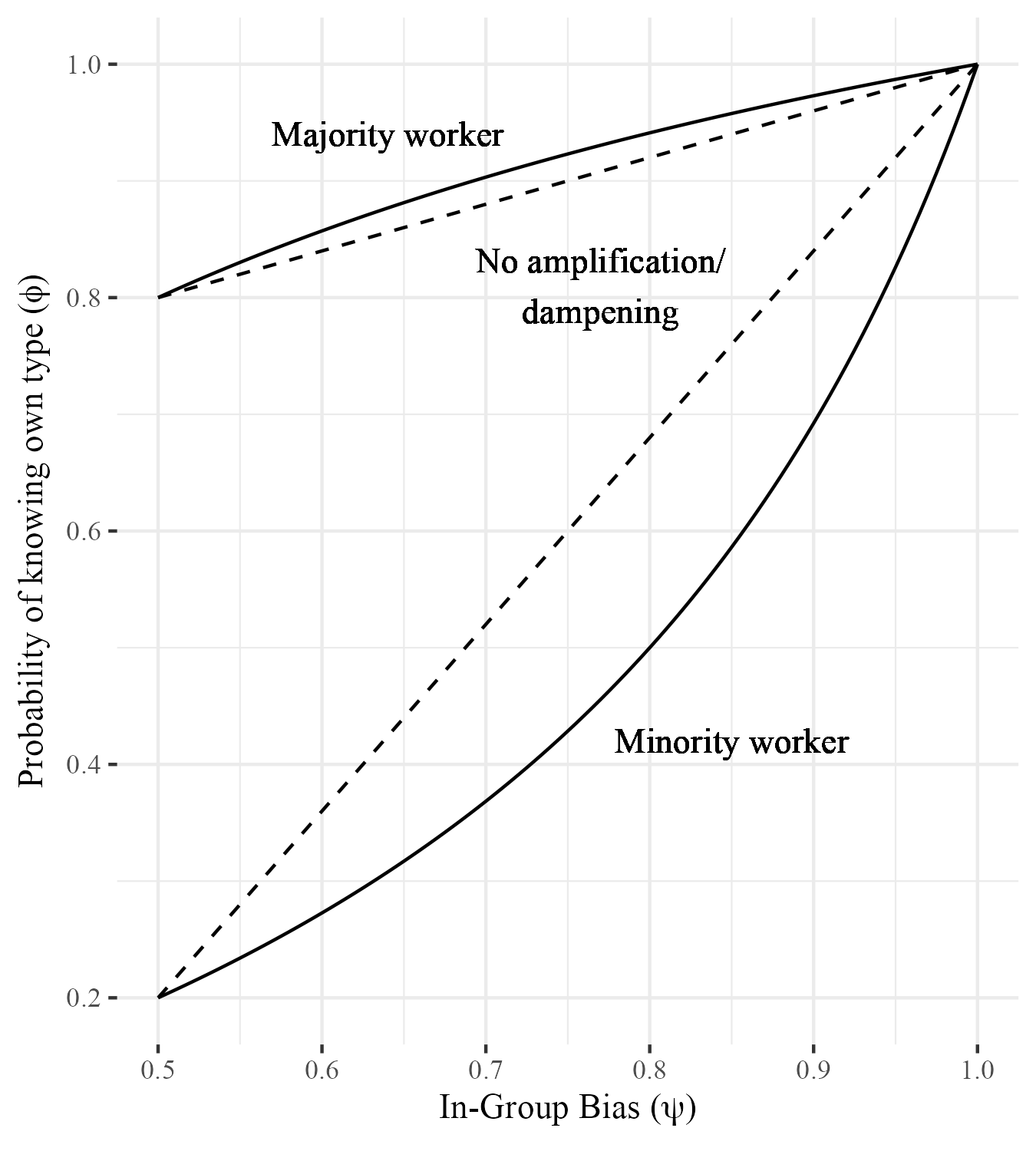}\\
\centering \footnotesize \textit{Note:} In this chart, the majority share of the labor force is 0.8.

\label{fig:sensitivity}
\end{figure}

The relationship between this paper's parameterization of in-group bias and the incidence of social ties can be seen in Figure \ref{fig:sensitivity}.  Specifically, Figure \ref{fig:sensitivity} illustrates the sensitivity of same-type social ties ($\phi$) to the in-group bias parameter ($\psi$).  This relationship is based on the expression linking share of the labor force ($w$) to in-group bias described in Equation \ref{eq:biascombined} of Proposition 1:%
\begin{align*}
 \Pr \{\text{period-1 worker knows own maj/min type}\} =\phi=\frac{w \cdot \psi}{[w \cdot \psi] + [(1-w)(1-\psi)]}
\end{align*}
(The Simplified Numerical Example immediately following Proposition 1 in Section \ref{sec:propdisc} illustrates the logic of this expression.)  When $\psi=0.5$, there is no bias (that is, the probability of social ties with the same type = $w$); when $\psi=1$, there is full bias (that is, the probability of social ties with the same type = 1). The dashed line represents a linear scaling, in which there is no amplification/dampening effect for majority and minority workers. Though not included in the graph below, the specification used in this paper yields a linear relationship if both workers' groups occupy 50\% of the labor force (that is, when social groups are the same size, there is no amplification/dampening effect).

The findings in this paper related to the unequal distribution of referral offers are robust to specifications for bias---for example, Equation \ref{eq:biascombined} from Proposition 1---where the relationship between the probability of knowing one’s own type  ($\phi$) and the bias ($\psi$) is more concave for majority workers than for minority workers.  In other words, it is robust to specifications where there is a comparative amplification effect on the incidence of same-type social ties for majority workers (as seen in Figure \ref{fig:sensitivity}).  This relationship captures the fact that an equal magnitude of bias has a disproportionately larger impact on majority workers than on minority workers. \par

\subsection{Why Another Homophily Parameterization is Not Used}
One might propose that instead of the parameterization I employ, one could simply define the homophily measures $\psi_{maj}$ and $\psi_{min}$ as the probability of referring another worker of the same type, conditional on having a social tie.\par

Under this alternative parameterization, for a given period-2 majority worker of high ability, the probability of receiving a referral from a specific period-1 majority worker would be $\frac{\tau_{maj} \alpha \psi_{maj}}{\delta N}$.  The numerator measures the probability that the referral is made to a period-2 majority high-ability worker, and the denominator is the probability that the offer is made to the specific period-2 worker from the pool of all available period-2 high-ability majority workers, of which there are $\delta N$. \par

There are $\delta N$ period-1 high-ability majority workers, so the expected number of referrals to a given period-2 worker of the same type is $\tau_{maj} \alpha \psi_{maj}$.  This is the expected value of $\delta N$ Bernoulli trials where each trial succeeds with probability $\frac{\tau_{maj} \alpha \psi_{maj}}{\delta N}$. \par

By similar logic, the expected number of referrals from period-1 high-ability minority workers to period-2 high-ability majority workers is:%

\begin{align*}
 \frac{(1 - \delta)\tau_{min} \alpha (1 - \psi_{min})}{\delta}
\end{align*}%
The expected total number of referrals made to any period-2 high-ability majority worker is the sum:
\begin{align}
 &\tau_{maj} \alpha \psi_{maj} + \frac{(1-\delta) \tau_{min} \alpha (1-\psi_{min})}{\delta}
\end{align}%
By similar logic, the expected number of offers made to any given period-2 high-ability minority worker is:
\begin{align}
 &\tau_{min} \alpha \psi_{min} + \frac{(\delta) \tau_{maj} \alpha (1-\psi_{maj})}{1-\delta}
\end{align}%
If we set $\tau_{maj} = \tau_{min}$ and $\psi_{maj}=\psi_{min}$, it is easy to see that the expression in (6) is greater than (7) when $\delta > \frac{1}{2}$.  However, this alternative parameterization of homophily suffers serious shortcomings.  To see why, consider the following two scenarios: (A) the majority group occupies 99\% of the population and $\psi_{maj}=\psi_{min}=0.99$, which would mean under this alternative parameterization that workers refer the same type 99\% of the time; and (B) the majority group only occupies 51\% of the population and $\psi_{maj}=\psi_{min}=0.99$.  

Despite capturing the likelihood of same-type referrals in both scenarios (A) and (B), $\psi$ \textit{under this alternative parameterization does not meaningfully reflect the level of bias exhibited by workers in forming same-type social ties}.  For example, a 99\% likelihood of referring the same type when one is the overwhelming majority has a significantly different interpretation than a 99\% likelihood of referring the same type when one is barely the majority, let alone when one is the minority.  As such, $\psi_{maj}=\psi_{min}$ does not reflect equal homophily between groups under this alternative model.  Thus, a homophily parameterization that does not explicitly incorporate the relative sizes of the in-group and out-group in determining the likelihood of same-type social ties suffers severe shortcomings.  The parameterization I use in this paper avoids these shortcomings through the relationship of $\psi$ with $\phi$, and represents a far superior measure.

\subsection{Model Implications in Historical Context}

Recall that the predicted inequality between majority and minority workers is based on several assumptions, which include: (1) majority and minority workers have no labor market disparities in the initial time period; (2) the only distinguishable difference between groups is relative size (that is, ability, network density, and in-group biases are all equivalent); (3) workers are more likely to know others with similar characteristics; and (4) there is no psychological prejudice.  These assumptions are critical when considering historical examples where minority workers enjoy \textit{greater} welfare than majority workers (for example, white South Africans), or when particular demographic groups who comprise a majority of a local labor market face worse outcomes (for example, black workers in a variety of U.S. metropolitan areas).  These cases do not undermine the accuracy of the model, not only because their circumstances clearly violate the model's assumptions (for example, that there is full equality between groups in the initial time period), but also because these cases intimately involve the distorting influence or legacy of psychological prejudice, which the employment model explicitly and intentionally does not incorporate. \par

\subsection{Model Implications in the University Context}
Social network discrimination matters in settings in which opportunities are distributed via social networks---such as within workplaces, in professional schools, or on college campuses.  For instance, college admissions---which involves admissions officers manufacturing an immersive academic and social community for years during a formative stage of life---inherently creates for many students the foundational network of lifelong personal and professional social ties.  As this paper's model suggests, race-neutral or ``colorblind'' policies in the construction of such social networks may exacerbate racial disparities.  Disproportionately fewer social ties may be formed by the minority group, all else equal, potentially translating to comparatively lower future professional opportunities, income, and welfare. \par

Furthermore, social network discrimination can lead to minorities suffering disadvantages in economic and social opportunities in ways that extend beyond employment referrals.  Opportunities, even while still in school, are often distributed based on social networks.  For example, many opportunities in school are distributed via informal information channels.  If minorities are harmed by social network discrimination, then all else equal, the dissemination of critical information on personal, educational, and career opportunities that is a hallmark of the college experience might be unequally distributed, contributing to disparities in outcomes between majority and minority groups.

\section{Calibration} \label{sec:modelcalibration}

This section calibrates the model to assess, in one application, the magnitudes of model parameters---such as majority and minority type in-group bias and network density---and to estimate a lower bound for the magnitude of inequality arising from social network discrimination.  The setting for this calibration exercise is the Public Use data from the National Longitudinal Study of Adolescent to Adult Health, 1994--2008 \parencite{harris2022}.  This data consists of a nationally representative sample of 7th- through 12th-grade American adolescents during the 1994--1995 school year.  In the friendship section of the in-school questionnaire, which was administered to over 90,000 students attending 145 schools in 80 communities, the respondents were asked to nominate up to five male and five female friends from the collection of all students enrolled.  This information was used to construct social network parameters, which are adapted below to estimate the parameters in this paper's model.  Further details about the overall sample and design of the study are provided in \textcite{resnick1997}. I make some simplifying assumptions for the purpose of this calibration, such as assuming that the social network parameters of the respondents do not change from childhood through their eventual entry into the labor market.

\subsection{Calibrated Social Network Parameters}
For the purpose of this calibration exercise, white respondents represent the majority, and black respondents represent the minority.  Here, I define some of the terms used below in the calibration data: \textit{ego} means respondent; \textit{alter} means the student in the same school as ego who is eligible to be nominated as a friend; \textit{node} means a unique member of a network; and \textit{ego's send-network} means the ego and the set of alters nominated by the ego as friends. Social network parameters are calculated as follows:\par

\textit{Majority share ($\delta$)---}This parameter is calculated by taking the total count of white respondents in the calibration data and dividing it by the total count of white respondents plus black respondents. Based on this definition, $\delta = 0.70$.

\textit{Ability in-group bias ($\alpha$)---}I vary this parameter between 0 and 1.\footnote{As illustrated in the Proof for Proposition 1B in the Appendix, the ability in-group bias does not impact the magnitude of social network parameters that is needed to achieve parity between majority and minority workers in the distribution of referral offers.}

\textit{Network density ($\tau_{maj}$ and $\tau_{min}$)---}\textcite{marsden1987} estimates the network density for whites at 0.61 and blacks at 0.63.\footnote{This may represent a conservative difference between $\tau_{maj}$ and $\tau_{min}$.  \textcite{marsden1987} also estimates the white population has significantly larger network sizes (mean size 3.1) than the black population (mean size 2.25).  This means that the average white worker has a greater probability of having a social tie (and thus receiving a referral).  Since our model sets a maximum of one social tie per period-1 worker, to account for this real-life difference in network sizes between races one might widen the gap in the network density parameters $\tau_{maj}$ and $\tau_{min}$ (which has not been done for the purposes of this calibration).} 

\textit{Type in-group bias ($\psi_{maj}$ and $\psi_{min}$)---}These parameters are calculated based on the calibration data parameter ``ego-network heterogeneity measure for race.''  The ego-network heterogeneity assesses the heterogeneity of the respondent's network with respect to race.  The formula used to calculate the ego-network heterogeneity for respondent $i$ is:%
\begin{align*}
& HETEROGENEITY_{iR} = 1 - \left[ \sum^n_{k=1} \left( \frac{R_k}{d}\right)^2 \right]  \hspace{.1cm} \text{,}\\
&\text{where:}\\
& R = \text{the race attribute}\\
& R_k = \text{the number of nodes with race } k \text{ in the ego network}\\
& d = \text{the number of nodes in the ego network with valid data on } R\\
& n = \text{the total number of races of } R \text{ represented in the ego network}
\end{align*}%

If all members of the respondent's network who have valid data on race share the same race, $HETEROGENEITY_{iR}=0$.\footnote{$HETEROGENEITY_{iR}$ is missing if respondent is the only member of their underlying network, or if all members of the respondent's network (including themselves) have missing data on race.}  I perform a simple transformation of $HETEROGENEITY_{iR}$ and take the mean to estimate the parameters $\phi_{maj}$ and $\phi_{min}$ (which represents the probability that a period-1 worker knows own maj/min type):%
\begin{align*}
& \phi_{maj} = mean \left(1 - HETEROGENEITY_{iR\text{ for i}\in\text{white respondents}}\right) \hspace{.1cm} \text{, and}\\
& \phi_{min} = mean \left(1 - HETEROGENEITY_{iR\text{ for i}\in\text{black respondents}} \right)
\end{align*}%
This gives $\phi_{maj}=0.74$ and $\phi_{min}=0.66$.  Then, I solve for $\psi_{maj}$ and $\psi_{min}$ by adapting Equation \ref{eq:biascombined} from Proposition 1 (which produces expressions that correspond to Equations \ref{eq:biasmaj} and \ref{eq:biasmin}):%
\begin{align*}
 &\Pr \{\text{white respondent refers white person}\} =\phi_{maj}=\frac{\delta \cdot \psi_{maj}}{[\delta \cdot \psi_{maj}] + [(1-\delta)(1-\psi_{maj})]} \\
 &\Pr \{\text{black respondent refers black person}\} =\phi_{min}=\frac{(1 - \delta) \cdot \psi_{min}}{[(1 - \delta) \cdot \psi_{min}] + [\delta \cdot (1-\psi_{min})]}
\end{align*}%
Substituting parameter values and solving gives $\psi_{maj} = 0.55$ and $\psi_{min}=0.82$.

\subsection{Inequality Arising From Social Network Discrimination}
Let us now take the calibrated model parameters, calculated above from a nationally representative sample of social networks among white and black respondents, to estimate the degree of economic inequality deriving purely from social network discrimination.  It is worth noting that the inequality predicted in this subsection is based on the counterfactual world introduced by the model: namely, that there is equal average ability, education, and initial employment between white and black workers.  I assume that half of white respondents and half of black respondents are high-ability, and half of both racial groups are low-ability.  Hence, the assumption moving forward is that the only difference between racial groups is in their social network parameters, not in the distribution of any attribute correlated with productivity or ability. \par

I have already calculated all relevant unknown parameters of our model in the previous subsection: $\delta = 0.70$, $\tau_{maj} = 0.61$, $\tau_{min} = 0.63$, $\psi_{maj} = 0.55$, and $\psi_{min}=0.82$.  

Next, I seek to determine the welfare gap---the difference in expected wage---between high-ability white and black respondents.  Let $H_{maj}$ denote high-ability majority worker, $H_{min}$ denote high-ability minority worker, $L_{maj}$ denote low-ability majority worker, and $L_{min}$ denote low-ability minority worker.  The steps I take to calculate the expected wage gap between high-ability black and white workers are as follows: 
\begin{enumerate}
    \item Estimate the likelihood that a high-ability white worker accepts a referral wage, relative to the likelihood a high-ability black worker accepts a referral wage;
    \item Estimate the market wage and the expected referral wage for both racial groups;
    \item Estimate the expected welfare gap between black and white workers by summing the market wage $w_{M2}$ and expected referral wage $E(w_{R_k})$, weighted by the likelihood of the worker’s racial group gaining employment through the general market or through the referral market, respectively.
\end{enumerate}
Computational details on these steps can be found in Appendix \ref{sec:appendixb}.

\begin{figure}
\captionsetup{justification=centering}
\caption{Lower-Bound Welfare Gap from Social Network Discrimination \\ (Black vs. White Workers)}
\centering
    \includegraphics{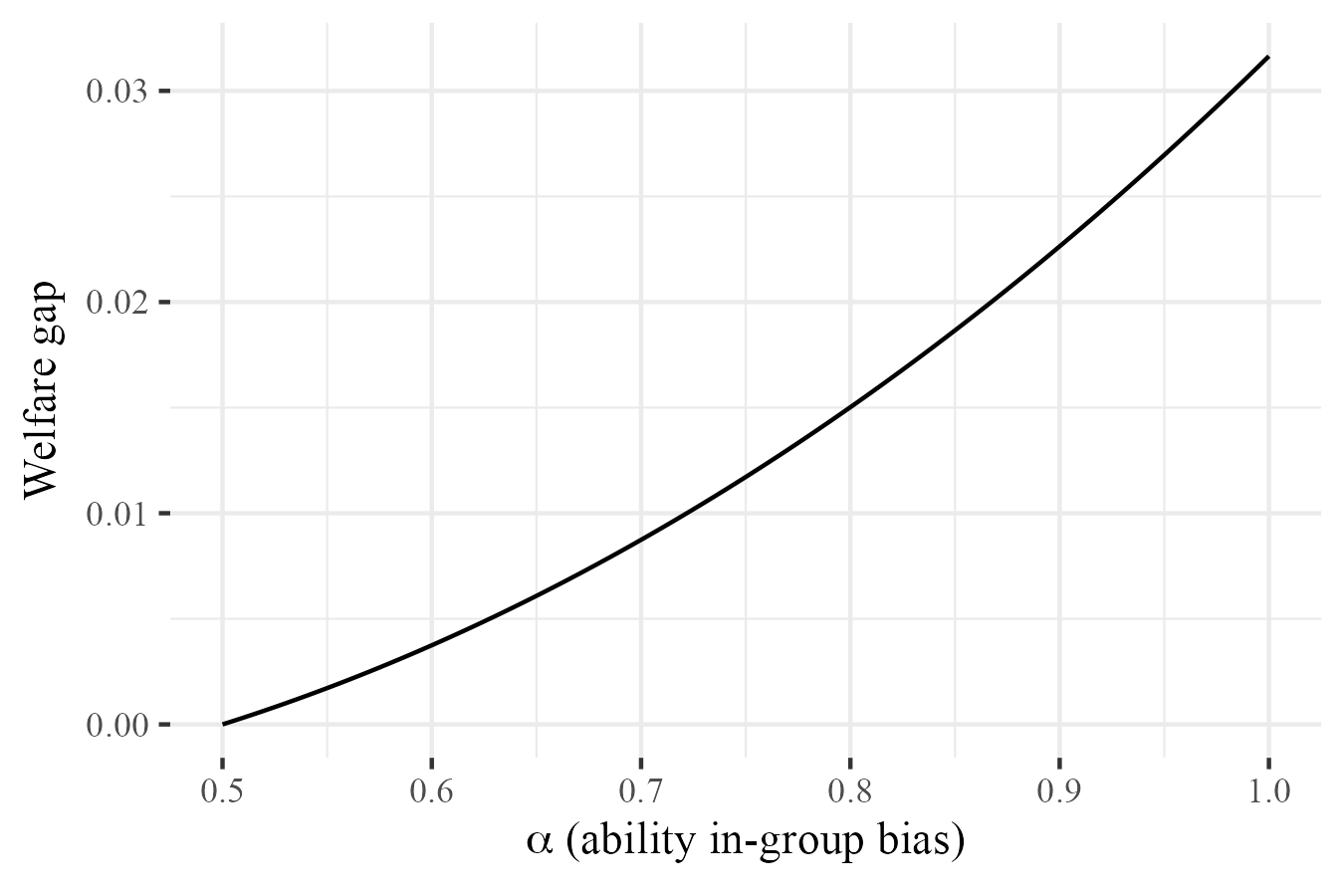}
\label{fig:calibration}
\end{figure}

After completing Steps 1 through 3, substituting calibration parameter values gives a welfare gap---which is the difference in expected wages in this model.  As I allow ability in-group bias to vary, the welfare gap varies in Figure \ref{fig:calibration} from 0 to 3.2 percent between racial groups, with black workers being disadvantaged compared to white workers.  This welfare gap is driven by two factors: (1) the disproportionately high likelihood that black workers will be hired through the (non-referral) general market, which pays the lowest wage on the equilibrium wage distribution; and (2) the disproportionately high volume of referrals white workers receive, which increases the maximum referral wage white workers are able to eventually accept compared to black workers.%

It is important to remember that the estimated welfare gap between black and white workers is based on a key simplifying assumption of the model: namely, that both the majority and minority groups begin the first period in a state of equality.  In other words, the maximum welfare gap in Figure \ref{fig:calibration} of 3.2 percent is what the model estimates over time, given an initial state of equality between racial groups.  Yet in the U.S. context, initial equality between racial groups did not exist; black workers not only may be negatively impacted today by persisting social network discrimination (as suggested above), but also they have been harmed historically by the remnants of past discrimination (which is outside the scope of this paper's analysis).  Historical discrimination would plausibly exacerbate the inequalities in referral opportunities between black and white workers---even conditional on the same magnitude of the underlying social network parameters.  As such, the estimated welfare gap from the calibration exercise may represent a lower bound of the true extent to which social network discrimination contributes to racial inequality in the U.S. context.

\section{Conclusion} \label{sec:modelconclusion}

The findings in this paper have significant theoretical, practical, and legal implications. \par

\textbf{Theoretical.} This paper terms the phenomenon called social network discrimination and reveals shortcomings of referral-based systems and related hiring practices from a racial justice perspective.  Despite initial equality in ability, employment, wages, and network structure---minorities receive fewer jobs through referral and lower expected wages simply because their social group is smaller.  This discriminatory outcome falls outside the dominant economics discrimination models---taste-based and statistical.  I calibrate the model and estimate the lower-bound welfare gap caused by social network discrimination is up to 3.2 percent, disadvantaging black workers.  This paper isolates a potential underlying mechanism for inequality, adding to the understanding of labor market disparities that have been widely studied across the social sciences. \par

\textbf{Practical.} By identifying social network discrimination, this paper introduces further important practical implications.  Research has shown that theories of discrimination and the language used in communicating them can rationalize or challenge stereotypes, as well as shape the attitudes and behaviors of influential members of society \parencite{tilcsik2021}.  Much popular U.S. debate already derides race-conscious policies as a deviation from merit or fairness; a wider awareness of social network discrimination could challenge these perceptions, highlighting that the root of persistent racial disparities may be more complex than previously understood. \par

Social network discrimination is applicable to a wide variety of real-world contexts beyond simply referrals for job opportunities.  Opportunities within a school can be unfairly distributed due to social network effects; the findings from this paper can be reasonably generalized to other settings in which opportunities are distributed based on informal information channels.   For example, the basic dissemination of personal, educational, and career guidance and wisdom that is a hallmark of the college experience can similarly be unfairly distributed due to social network discrimination.  Some peers may have greater access to information simply because they belong to a larger social group. \par

Important social ties are formed in postsecondary schooling (for example, colleges and universities), in professional schools (for example, business schools, law schools, and medical schools), and in workplaces.  This paper suggests there are inherent social network advantages associated with belonging to a larger demographic group, all else equal, since the formation of such ties is influenced by social phenomena like homophily.  In short, social network discrimination may distort the allocation of personal, educational, and professional opportunities in significant ways that foster racial disparities---an important insight that not only may help explain substantial inequalities that persist between demographic groups today, but also may help inform efforts to promote diversity and equitable access to opportunity in the future.\par

\textbf{Legal.} This paper disproves a persisting argument that race-blind or ``colorblind'' policies are inherently merit-enhancing, introducing a new rationale for race-conscious policy.  In this paper's model, employers cannot identify which workers are majority and which ones are minority---that is, employers operate fully under color-blind hiring policies---yet disparities still arise between demographic groups.  In enforcing Title VII of the Civil Rights Act, the EEOC prohibits employers from using neutral employment policies and practices that have a disproportionately negative impact on applicants or employees of a particular color, race, age, sex, national origin, or disability status.  Referral-based hiring may constitute, in some contexts, such a policy with disproportionately negative impact.  Furthermore, social network discrimination is relevant for ongoing legal (and popular) debates surrounding affirmative action---both in the workplace and in higher education \parencite{okafor2024}.\footnote{\textcite{okafor2024} introduces the concept of social network discrimination to legal scholarship and begins to explore how it relates to the jurisprudence surrounding affirmative action.}  To date, the Court has not meaningfully considered remedying social network discrimination as a rationale for race-conscious policy.  Lastly, this paper has implications even moving beyond race, challenging the merits of the anticlassification interpretation of the Equal Protection Clause of the Fourteenth Amendment.\footnote{\textcite{balkin2003} states that, ``roughly speaking, [the anticlassification principle] holds that the government may not classify people either overtly or surreptitiously on the basis of a forbidden category.''  Forbidden categories could be interpreted to align with the federal protected classes, which include (among others): race, color, religion or creed, national origin or ancestry, sex (including gender, pregnancy, sexual orientation, and gender identity), age, and physical or mental disability.}  As mentioned earlier, the majority/minority characterization in this paper is tied to race due to social science research that finds that race creates the greatest social divide.  Yet it is entirely possible that in certain environments, homophily along other personal attributes (for example, gender or sexual orientation) may also be sufficiently meaningful, if not more dominant than racial categorizations.  Thus, corrective policies that explicitly factor in such personal attributes may be needed to achieve true equality of opportunity in different settings.  This policy stance conflicts with the anticlassification principle that currently dominates Supreme Court equal protection jurisprudence. \par

I leave the specific contours of all these legal arguments for future research.  At a bare minimum, simply recognizing the distorting influences of social network discrimination can better inform how we as a society approach matters of race and justice.

\newpage

\pagebreak

\renewcommand{\thesection}{\Alph{section}}
\setcounter{section}{0}

\section{Appendix: Model Proofs} 
\phantomsection
\label{sec:appendixa}
\addcontentsline{toc}{section}{Appendix A}

\setcounter{proposition}{1}
\setcounter{theorem}{0}

\setcounter{theorem}{0}

\begin{lemma}
In equilibrium, referral wage offers lie within the interval between $w_{M2}$ and a maximum referral wage offer $\overline{w}_{R}$; hence, \emph{$w_{R} \in [w_{M2}$, $\overline{w}_{R}$]}.  The density of the referral wage offer distribution is positive across this entire range.
\end{lemma}
\begin{proof}
I begin by arguing the existence and uniqueness of an equilibrium.\footnote{The first two paragraphs of this proof are modified from \textcite{bolte2020}.  The last paragraph in this proof is modified from \textcite{montgomery1991}.}  To do so, I begin by proving the existence of a fixed point.  Let the correspondence $C$ be
\begin{align*}
    C(\widetilde{a})=\{\text{max}[0, E_{\widetilde{a},r}[v_q|q\in \text{market}]]|r \in [0,1]\}
\end{align*}
This correspondence builds on Equation \ref{eq:equilibrium}, the expression that characterizes the equilibrium threshold (with ``equilibrium threshold'' referring to a fixed point of the expression).  Varying over $r$ impacts the expected productivity in the market at atoms in the distribution.  The existence of an equilibrium threshold can be seen by the fact that $C$ is convex-valued, compact-valued, and upper-hemicontinuous---and is bounded above (by some $\overline{w}_R$ as described later in this proof) and below (by 0, both the minimum possible productivity and option value for any period-2 worker).  Therefore, there is a fixed point (by Kakutani's Theorem).  Firms can mix strategies arbitrarily at a fixed point; this comes from the fact that if the equilibrium threshold equals the expected productivity in the market, $\widetilde{a}=E_{\widetilde{a},r}[v_q|q\in \text{market}]$, then hiring a given worker or leaving them in the general market does not change the average productivity in the market.  This is because, when the equilibrium threshold equals the expected productivity in the market, the productivity of the referred period-2 worker and the expected productivity in the market are exactly equal.  

\par

Here I prove the uniqueness of the fixed point.  Suppose instead that there were two separate equilibrium thresholds, $\widetilde{a}<\widetilde{a}'$.  In this construction, $\widetilde{a}' > w_{M2}$, as otherwise it must be that $\widetilde{a}=\widetilde{a}'=w_{M2}$, since the correspondence $C$ is bounded below by $w_{M2}$.  The general market for threshold $\widetilde{a}'$ is comprised of period-2 workers who did not receive any referral offers and period-2 workers referred by period-1 workers of productivity between $\widetilde{a}$ and $\widetilde{a}'$ (and potentially some workers with productivity exactly at the threshold $\widetilde{a}'$, depending on how firms are mixing their strategies).  The expected productivity of period-2 workers who do not receive any referrals is less than or equal to $\widetilde{a}$, since $\widetilde{a}$ is an equilibrium threshold.  The expected productivity in the market if the threshold is $\widetilde{a}'$ can be calculated by taking the mean productivity of this group with the expected productivity of period-2 workers referred by period-1 workers of productivity between $\widetilde{a}$ and $\widetilde{a}'$.  However, this setting cannot result in an expected productivity of $\widetilde{a}'$, as $\widetilde{a} <\widetilde{a}'$.

Regarding wage dispersion: the market wage ($w_{M2}$) must coincide with the bottom of the referral wage distribution, as any referral offer below the market wage will necessarily be rejected by workers in favor of going to the general market.  At the maximum referral wage offer (denoted $\overline{w}_{R}$ and derived in Appendix Equation \ref{eq:maxreferral}), the probability a worker accepts the referral offer is 1 (as there are no better wage offers in either the referral market or general market).  Hence, firms will not offer a referral wage above this amount as it will necessarily reduce profits.  Proposition 2.2 in \textcite{butters_equilibrium_1977} Butters (1977)\footnote{GR Butters, \textit{Equilibrium Distribution of Prices and Advertising in Oligopoly}, 44 \textsc{Review of Economic Studies}, 465, 470 (1977).} proves that there are no gaps in the wage distribution.  If there were a gap between some $w_1$ and $w_2$, then a firm offering the higher wage could reduce its offer by $\epsilon$ without reducing the probability its offer is accepted, thereby increasing its profits.
\end{proof}

\begin{lemma}
A firm will attempt to hire through referral if and only if it employs a high-ability worker in period 1.
\end{lemma} 

\begin{proof}
Firms employing high-ability workers in period 1 will make referral offers for two reasons.\footnote{This proof for Lemma 2 is modified from \textcite{montgomery1991}.}  First, hiring through the market yields zero expected profit (given the assumption of free entry of firms).  Second, an accepted referral offer yields constant positive profit over the range of the referral offer wage distribution $[w_{M2} ,\overline{w}_{R}]$.  (If it did not yield constant positive profits, it would be impossible to maintain equilibrium wage dispersion; the profit-maximizing firms would offer only a subset of the distribution---that is, those wages that maximized profits.)  An offer below $w_{M2}$ will never be accepted---as workers will be able to get higher wages by getting hired in the general period-2 market---while an offer above $\overline{w}_{R}$ increases the wage without increasing the probability of attracting a worker (since $\overline{w}_{R}$ is the top of the referral offer wage distribution and so has a 100\% chance of being accepted).  To finish this proof, I will show that firms employing low-ability workers in period 1 will hire through the market (that is, not rely on referrals). \par

If a firm employing a low-ability worker did deviate from this Lemma and made a referral offer $w_{R}$, its expected profit (denoted $E\prod _{L}(w_{R})$) would be represented by a slight modification to the expression for the expected profit for a firm employing a high-ability worker (denoted $E\prod _{H}(w_{R})$ and derived in Equation \ref{eq:expectedprofit} of Proposition 3).  The expression for $E\prod _{L}(w_{R})$ differs from that of $E\prod _{H}(w_{R})$ in that the incidences of $\alpha$ in the first $p^{HMAJ}$, $p^{HMIN}$, $p^{LMAJ}$, and $p^{LMIN}$ terms are replaced with ($1 - \alpha$), and vice versa. \par

In this scenario, as long as $\alpha>\frac{1}{2}$, $E\prod _{L}(w_{R})<0$.  Hence, expected profits are less than those from hiring in the general market, which equals zero due to free entry of firms.  The lemma is thus proved: a firm employing a low-ability worker in period-1 prefers to hire through the market, maximizing expected profit. \end{proof}

\begin{lemma}
The period-1 market wage is greater than the expected period-1 productivity.
\end{lemma} 

\begin{proof}
Firms hiring in the period-1 market earn an expected period-2 profit equal to the probability of hiring a high-ability period-1 worker times the expected profit (denoted $c$) from a referral.\footnote{This proof for Lemma 3 is modified from \textcite{montgomery1991}.}  Due to free entry, the wage is driven above expected period-1 productivity: %
\begin{align*}
    & w_{M1}(\alpha ,\delta  ,\tau _{maj} ,\tau _{min} ,\psi _{maj} ,\psi _{min}) \\
    &=\frac{1}{2} +\frac{1}{2}c(\alpha ,\delta ,\tau _{maj} ,\tau _{min} ,\psi _{maj} ,\psi _{min}) \\
    & =\frac{1}{2}[1 +c(\alpha ,\delta ,\tau _{maj} ,\tau _{min} ,\psi _{maj} ,\psi _{min})]
\end{align*}%
The expression for $c$ is derived in Proposition 3.  Given comparative-statics results on $c$, $w_{M1}$ is increasing in $\tau _{maj}$ and $\tau _{min}$. When both $\tau _{maj}=\tau _{min}$ and $\psi _{maj}=\psi _{min}$, $w_{M1}$ is decreasing in $\delta$.
\end{proof}

\begin{subprop}
In an environment with equal magnitude of majority/minority network parameters ($\tau_{maj}=\tau_{min}$ and $\psi_{maj}=\psi_{min}$), the probability that, among referral offers, a minority worker is referred is lower than their share of the labor force.  Conversely, the probability that, among referral offers, a majority worker is referred is larger than their share of the labor force.
\end{subprop}
\begin{subprop}
The inequality in the distribution of referral job offers can be eliminated by minority workers having a sufficiently higher type in-group bias ($\psi_{min}$).
\end{subprop} 

\begin{proof}
\textit{[These proofs for Propositions 1A and 1B are modified from \textcite{montgomery1991}.]}  Let us first consider a given high-ability period-2 worker ($H$).  Since all referral wage offers are above the period-2 market wage, the probability that $H$ would accept a referral wage offer $w_{Ri}$ from firm $i$ can be expressed:%
\begin{align*}
    \Pr \{\text{$H$ accepts $w_{Ri}$}\} = \Pr \{\text{$H$ receives no higher offer $w_{Rj}\  \forall  $ firm $j \neq i$}\}
\end{align*}
\noindent Since referral offers are allocated independently, %
\begin{align*}
    \Pr \{\text{$H$ accepts $w_{Ri}$}\} &= \prod _{j \neq i}\Pr \{\text{$H$ receives no higher offer $w_{Rj}$}\} \\
        & =\prod _{j \neq i}[\ensuremath{\operatorname*{}}1 -\Pr \{\text{$H$ receives an offer $w_{Rj} >w_{Ri}$}\}]
\end{align*}
\noindent The probability that firm $j$ offers a wage $w_{Rj} >w_{Ri}$ to $H$ is the product of two independent probabilities: %
\begin{align*}
  \Pr \{\text{$H$ receives an offer $w_{Rj} >w_{Ri}$}\} =\Pr \{\text{firm $j$ makes offer to $H$}\} \cdot \Pr \{w_{Rj} >w_{Ri}\}  
\end{align*}
If $2N$ workers were in period-1, free entry implies that $N$ firms employ high-ability workers.  Now I will analyze both parts of the expression from the perspective of a high-ability majority worker ($H_{maj}$) and high-ability minority worker ($H_{min}$).  

The probability that firm $j$ offers a referral to $H_{maj}$ is a weighted average of whether firm $j$ hired a majority or minority worker in period 1.  Denote $\phi _{maj}$ as the probability a majority worker knows another majority worker, and $\phi _{min}$ as the probability a minority worker knows another minority worker, where:%
\begin{align*}
    \phi _{maj} =\frac{\left (\delta  \cdot \psi _{maj}\right )}{\left (\delta  \cdot \psi _{maj}\right ) +[(1 -\delta ) \cdot (1 -\psi _{maj})]} \hspace{.1cm} \text{, and}\\\\
    \phi _{min} =\frac{(1 -\delta ) \cdot \psi _{min}}{[(1 -\delta ) \cdot \psi _{min}] +[\delta  \cdot (1-\ \psi _{min})]} \hspace{1cm}
\end{align*}
Then: %
\renewcommand{\theequation}{A\arabic{equation}}
\setcounter{equation}{0}
\begin{align} \label{eq:proboffer}
\begin{split}
     &\Pr \{\text{firm $j$ makes offer to $H_{maj}$}\} = \delta \left (\frac{\alpha \tau _{maj} \phi _{maj}}{N}\right ) +(1 -\delta )\genfrac{(}{)}{}{}{\alpha \tau _{min}(1 -\phi _{min})}{N} 
\end{split}
\end{align}
\noindent Likewise, for a period-2 minority high-ability worker:
\begin{align*}
    \Pr \{\text{firm $j$ makes offer to $H_{min}$}\} = \delta \left (\frac{\alpha \tau _{maj}(1 -\phi _{maj})}{N}\right ) +(1 -\delta )\genfrac{(}{)}{}{}{\alpha \tau _{min} \phi _{min}}{N}
\end{align*}
\noindent Based on these expressions, for minority workers to have a proportional chance of receiving job offers through referral, the following must hold:%
\begin{equation*}
\begin{gathered}
     \Pr \{\text{firms make offer to $H_{maj}$}\} \propto \Pr \{\text{firms make offer to $H_{min}$}\}\\\\
     \text{only when}\\\\
     (1 -\delta ) \cdot \Pr \{\text{firms make offer to $H_{maj}$}\} = \delta \cdot \Pr \{\text{firms make offer to $H_{min}$}\}\\\\
     \text{or}\\\\
     (1 -\delta )\left [\delta \tau _{maj} \phi _{maj} +(1 -\delta )\tau _{min}(1 -\phi _{min})\right ] =\delta \left [\delta \tau _{maj} (1 -\phi _{maj}) +(1 -\delta )\tau _{min}(\phi _{min})\right ]
\end{gathered}
\end{equation*}

Given $\tau_{maj}=\tau_{min}$ and $\psi_{maj}=\psi_{min}$, analysis shows the left-hand side of the equation is always larger than the right-hand side of the equation, which means majority workers receive a disproportionately large volume of referral offers.  Both the minority network density required to compensate for the inequality (denoted $\tau _{min}^{=}$) and the minority type in-group bias required to compensate (denoted $\psi _{min}^{=}$) increase in $\tau _{maj}$, $\psi _{maj}$, and $\delta$.  The greater the probability of majority workers having social ties (or the greater the degree of their type in-group bias or likelihood of possessing social ties), the greater minority workers' compensating parameters ($\tau _{min}^{=}$ or $\psi _{min}^{=}$) must be to achieve a proportional amount of all job offers through referrals. \par

Though higher minority network density and type in-group bias can both reduce the disproportionality in the distribution of job offers through referral, analysis shows that between these two parameters only $\psi _{min}^{=}$ is attainable (that is, below 1) under all possible combinations of social network parameters.
\end{proof}

\begin{proposition}
In an environment with equal magnitude of majority/minority network parameters ($\tau_{maj}=\tau_{min}$ and $\psi_{maj}=\psi_{min}$), the period-2 market wage ($w_{M2}$) decreases as majority workers occupy a greater fraction of the labor force.  $w_{M2}$ also decreases in the ability in-group bias $\alpha$.
\end{proposition}

\begin{proof}
\textit{[This proof for Proposition 2 is modified from \textcite{montgomery1991}.]}  First I derive an expression for period-2 market wage.  I build on the analysis from Proposition 1.  If there were $2N$ workers in period 1, free entry implies that $N$ firms employ high-ability workers.  If firms select their referral wage offer by randomizing over the equilibrium wage distribution $F(\bullet )$ (to be derived below), %
\begin{align*}
   \Pr \{\text{$H_{maj}$ receives an offer $w_{Rj} >w_{Ri}$}\} = \Pr \{\text{firm $j$ makes offer to $H_{maj}$}\} \cdot [1 -F(w_{Ri})]\\
\end{align*}
\noindent for all firms $j$ who employ a high-ability worker in period 1.  I have already shown that:%
\begin{align*}
    \Pr \{\text{$H$ accepts $w_{Ri}$}\} &= \prod _{j \neq i}\Pr \{\text{$H$ receives no higher offer $w_{Rj}$}\} \\
        & =\prod _{j \neq i}[\ensuremath{\operatorname*{}}1 -\Pr \{\text{$H$ receives an offer $w_{Rj} >w_{Ri}$}\}]
\end{align*}
\noindent Substitution yields: %
\begin{align*}
    &\Pr \{\text{$H_{maj}$ accepts $w_{Ri}$}\} \\ 
    &=\{1 -[\frac{1}{N}(\delta \alpha \tau _{maj} \phi _{maj} + (1 -\delta) \alpha \tau _{min}(1 -\phi _{min}))] \cdot [1 -F(w_{Ri})]\}^{N -1}
\end{align*}
\noindent Since the model assumes a large number of workers, as $N$ approaches $\infty $,%
\renewcommand{\theequation}{A\arabic{equation}}
\begin{align} \label{eq:probaccepts}
    \Pr \{\text{$H_{maj}$ accepts $w_{Ri}$}\} =\exp \{ -[\delta \alpha \tau _{maj} \phi _{maj} + (1 -\delta) \alpha \tau _{min}(1 -\phi _{min})][1 -F(w_{Ri})]\}
\end{align}
\noindent Details on this step can be found in \textcite{rapoport1963} and \textcite{montgomery1991}.  One can use similar steps to obtain the probability that firm $i$'s offer is accepted by a given high-ability majority worker ($H _{maj}$), low-ability majority worker($L _{maj}$), and low-ability minority worker($L _{min}$). \par

As high-ability workers tend to receive more offers, they are less likely to accept any given offer $w_{Ri} <\overline{w}_{R}$.  Since a period-2 worker finds employment through the market only if he receives no offers (or rejects all referral offers): %
\begin{align*}
    \Pr \{\text{market \textbar{} $H_{maj}$}\} =\Pr \{\text{$H_{maj}$ accept $w_{M2}$}\}
\end{align*}
The market wage coincides with the bottom of the referral wage distribution, $F(\bullet )$, because any referral wage below the market wage will be rejected by period-2 workers, to gain employment through the market.  Thus, given that $F(w_{M2}) =0$: %
\begin{align*}
    \Pr \{\text{market \textbar{} $H_{maj}$}\} = \exp \{ -[\delta \alpha \tau _{maj} \phi _{maj} + (1 -\delta) \alpha \tau _{min}(1 -\phi _{min})]\}
\end{align*}
\noindent I can derive similar expressions for $H_{min}$, $L_{maj}$, and $L_{min}$.  Let:

\begin{align} \label{eq:Hmaj}
\begin{split}
    e^{HMAJ} &=\exp \{ -[ \delta \alpha \tau _{maj} \phi _{maj} + (1 -\delta) \alpha \tau _{min}(1 -\phi _{min})]\} \\
    e^{HMIN} &=\exp \{ -[ \delta \alpha \tau _{maj}(1 - \phi _{maj}) + (1 -\delta) \alpha \tau _{min}\phi _{min}]\} \\
    e^{LMAJ} &=\exp \{ -[ \delta (1 - \alpha) \tau _{maj} \phi _{maj} + (1 -\delta) (1 - \alpha) \tau _{min}(1 -\phi _{min})]\} \\
    e^{LMIN} &=\exp \{ -[ \delta (1 - \alpha) \tau _{maj}(1 - \phi _{maj}) + (1 -\delta) (1 - \alpha) \tau _{min}\phi _{min}]\} 
\end{split}
\end{align}

I now use Bayes's rule to calculate the period-2 market wage: %
\begin{align} \label{eq:wm2}
\begin{split}
    & w_{M2}(\alpha, \delta , \tau _{maj}, \tau _{min}, \psi _{maj}, \psi _{min}) \\
    &= E(\text{productivity \textbar{} market}) \\
    &=\frac{\Pr(\text{market \textbar{} $H_{maj}$}) \cdot \Pr(H_{maj})+\Pr(\text{market \textbar{} $H_{min}$}) \cdot \Pr(H_{min})}{\Pr(\text{market \textbar{} $H$}) \cdot \Pr(H)+\Pr(\text{market \textbar{} $L$}) \cdot \Pr(L)} \\
    &=\frac{(e^{HMAJ} \cdot \delta) +(e^{HMIN} \cdot (1 -\delta))}{(e^{HMAJ} + e^{LMAJ}) \cdot \delta  +(e^{HMIN} +e^{LMIN}) \cdot (1 -\delta )}
\end{split}
\end{align}
Given $\alpha > \frac{1}{2}$ and both network densities ($\tau _{maj}$ and $\tau _{min}$) greater than zero,  $w_{M2}$ is always less than $\frac{1}{2}$, the average productivity of the population.  Analysis shows that $w_{M2}$ is decreasing in $\alpha $.  Furthermore, for all $\psi _{maj} =\psi _{min}$ and $\tau _{maj} =\tau _{min}$, $w_{M2}$ is also decreasing in $\delta$. 
\end{proof}

\begin{proposition}
In an environment with equal magnitude of majority/minority network parameters ($\tau_{maj}=\tau_{min}$ and $\psi_{maj}=\psi_{min}$), the referral wage and the welfare (that is, average expected wage) for minority workers are lower than for majority workers.
\end{proposition} 

\begin{proof}
\textit{[This proof for Proposition 3 is modified from \textcite{montgomery1991}.]}  Consider the expected period-2 profit earned by a firm employing a high-ability worker and setting a referral wage (recall the productivity of high-ability workers equals one, while that of low-ability workers equals zero): \par
$E\prod _{H}(w_{R})$ %
\begin{align*}
    & = \Pr \{\text{high-ability majority period-2 referral hired \textbar{} $w_{R}$}\} \cdot (1 -w_{R}) \\
    & \qquad + \Pr \{\text{high-ability minority period-2 referral hired \textbar{} $w_{R}$}\} \cdot (1 -w_{R}) \\
    & \qquad +\Pr \{\text{low-ability majority period-2 referral hired \textbar{} $w_{R}$}\} \cdot (-w_{R}) \\
    & \qquad +\Pr \{\text{low-ability minority period-2 referral hired \textbar{} $w_{R}$}\} \cdot (-w_{R})
\end{align*}

\noindent (If no referred worker is hired, perhaps because the period-1 worker possesses no social tie or because the referred acquaintance receives a higher referral offer from another firm, the firm hires through the period-2 general market and earns zero expected profit.) \par

The probability of hiring a high-ability majority period-2 referred worker is the product of two independent probabilities (substituting from Equations \ref{eq:proboffer} and \ref{eq:probaccepts} in Propositions 1 and 2):%
\begin{align*}
    &\Pr \{\text{high-ability period-2 majority referral hired \textbar{} $w_{R}$}\} \\
    &=\Pr \{\text{offer made to a high-ability majority referral}\} \cdot \Pr \{\text{$H_{maj}$ accepts $w_{R}$}\} \\
    &= \delta \alpha \tau _{maj} \phi _{maj}\ +(1 -\delta )\alpha \tau _{min}(1 -\phi _{min}) \\
    & \qquad \cdot \exp \{ -[\delta \alpha \tau _{maj} \phi _{maj} +(1 -\delta )\alpha \tau _{min}(1 -\phi _{min})][1 -F(w_{R})]\}
\end{align*}
\noindent Similar steps can be followed to derive the respective conditional probability for high-ability minority, low-ability majority, and low-ability minority workers.\par
Let:%

\renewcommand{\theequation}{A\arabic{equation}}
\begin{align} \label{eq:pmaj}
\begin{split}
        p^{HMAJ} &= \delta \alpha \tau _{maj} \phi _{maj} +(1 -\delta )\alpha \tau _{min}(1 -\phi _{min}) \\
        p^{HMIN} &= \delta \alpha \tau _{maj} (1 - \phi _{maj}) + (1 -\delta) \alpha \tau _{min} \phi _{min} \\
        p^{LMAJ} &= \delta (1 -\alpha) \tau _{maj} \phi _{maj} + (1 -\delta) (1 - \alpha) \tau _{min} (1 - \phi _{min}) \\
        p^{LMIN} &= \delta (1 - \alpha) \tau _{maj} (1 - \phi _{maj}) + (1 -\delta) (1 - \alpha) \tau _{min} \phi _{min}
        \end{split}
\end{align}

\noindent So, to simplify: \par
\hspace{1cm} $E\prod _{H}(w_{R})$ 
\renewcommand{\theequation}{A\arabic{equation}}
\begin{align} \label{eq:expectedprofit}
\begin{split}
    &=p^{HMAJ} \cdot \exp \{ -[p^{HMAJ}][1 -F(w_{Ri})]\} \cdot (1 -w_{R}) \\
    & \qquad  +p^{HMIN} \cdot \exp \{ -[p^{HMIN}][1 -F(w_{Ri})]\} \cdot (1 -w_{R}) \\
    & \qquad +p^{LMAJ} \cdot \exp \{ -[p^{LMAJ}][1 -F(w_{Ri})]\} \cdot (-w_{R}) \\
    & \qquad  +p^{LMIN} \cdot \exp \{ -[p^{LMIN}][1 -F(w_{Ri})]\} \cdot (-w_{R})
\end{split}
\end{align}
For wage dispersion to be maintained in equilibrium, firms must earn the same expected profit on each referral wage offered:%
\begin{align*}
    E\prod _{H}(w_{R}) =c\qquad  \forall w_{R} \in [w_{M2} ,\overline{w}_{R}]
\end{align*}
To calculate this profit constant, note that the firm could deviate from the specified strategy and offer a wage of $w_{M2}$; in this case, the referred worker accepts the firm's offer only if they receive no other higher referral offers. \par 

Recall that $F(w_{M2})=0$.  The firm's expected profit is therefore given by (using terms defined in Equations \ref{eq:Hmaj} and \ref{eq:pmaj}):%
\begin{align*}
    &E\prod _{H}(w_{M2}) \\
    &=(p^{HMAJ})(e^{HMAJ})(1 -w_{M2}) +(p^{HMIN})(e^{HMIN})(1 -w_{M2}) \\
    &\qquad  +( p^{LMAJ})(e^{LMAJ})(-w_{M2}) +(p^{LMIN})(e^{LMIN})(-w_{M2}) \\
    &=c
\end{align*}
\noindent Substituting for $w_{M2}$ (Equation \ref{eq:wm2}), I can determine $c(\alpha ,\delta  ,\tau _{maj} ,\tau _{min} ,\psi _{maj} ,\psi _{min})$.  Given $\alpha > \frac{1}{2}$, firms with high-ability workers who possess social ties earn positive expected profits.  Analysis shows that $c$ is increasing in $\alpha$, $\tau _{maj}$, and $\tau _{min}$.  When both $\psi_{maj}=\psi_{min}$ and $\tau_{maj}=\tau_{min}$, $c$ is decreasing in $\delta$.\par

Given the expression for $c(\alpha ,\delta  ,\tau _{maj} ,\tau _{min} ,\psi _{maj} ,\psi _{min})$, the equilibrium referral-offer distribution $F(\bullet )$ can be determined by setting $E\prod _{H}(w_{R})$ (Equation \ref{eq:expectedprofit}) equal to $c$ for all potential wage offers $w_{R}$. \par

Unfortunately, doing so does not result in a closed-form solution for $F(w_{R})$.  As \textcite{montgomery1991} also finds, the equilibrium referral-offer distribution $F(\bullet )$ can be interpreted as either: (1) each firm randomizes over the entire distribution; or (2) a fraction $f(w_{R})$ of firms offers each wage for certain. \par

From the second interpretation, one can denote these referral wages with $w_{Rk}$ and estimate the average referral wage received by a high-ability majority worker (denoted $E(w_{R _{Hmaj}})$) vs. a high-ability minority worker (denoted $E(w_{R _{Hmin}})$), for any given $\delta$, $\alpha$, $\tau _{maj}$, $\tau _{min}$, $\psi _{maj}$, and $\psi _{min}$.  Analysis shows that in an environment with equal magnitude of majority/minority network parameters, if $\alpha > \frac{1}{2}$ and $\delta  >\frac{1}{2}$, $E(w_{R _{Hmaj}}) >E(w_{R _{Hmin}})$. \par

Proposition 1 shows that, all else equal, minority workers receive a smaller proportion of jobs through referral than their fraction of the population.  As a result, minority workers more frequently gain employment through the market, receiving the (lower) market wage.  In this Proposition, I showed that even when offered a job through referral, minority workers have lower expected referral wages than majority workers.\end{proof}

To conclude, one can derive an expression for the maximum referral wage offered $\overline{w}_{R}$ (where $F(\overline{w}_{R}) =1$, by definition): \par
\renewcommand{\theequation}{A\arabic{equation}}
\begin{align} \label{eq:maxreferral}
    \overline{w}_{R}(\alpha ,\delta ,\tau _{maj} ,\tau _{min} ,\psi _{maj} ,\psi _{min}) =\frac{p^{HMAJ} + p^{HMIN} -c}{p^{HMAJ} + p^{HMIN} + p^{LMAJ} + p^{LMIN}}
\end{align}
A firm that offers a referral wage of $\overline{w}_{R}$ attracts a referred worker with probability 1 (conditional on its period-1 worker possessing a social tie).  The firm's expected profit, $c$, is thus equal to $p^{HMAJ} + p^{HMIN}-\overline{w}_{R}(p^{HMAJ} + p^{HMIN} + p^{LMAJ} + p^{LMIN})$.  $\overline{w}_{R}$ is increasing in $\alpha$, $\tau _{maj}$, and $\tau _{min}$.

\pagebreak

\section{Appendix: Calibration Steps} 
\phantomsection
\label{sec:appendixb}
\addcontentsline{toc}{section}{Appendix B}
\renewcommand{\thesubsection}{B\arabic{subsection}}

\subsection{Step 1: Estimate the relative likelihood between racial groups that a high-ability worker accepts a referral wage}

As Appendix \ref{sec:appendixa} explains, let $e^{X}$ denote the $Pr(X\text{ accepts the market wage})$, which can be calculated as follows:
\begin{align*}
\begin{split}
    e^{H_{maj}} &=\exp \{ -[ \delta \alpha \tau _{maj} \phi _{maj} + (1 -\delta) \alpha \tau _{min}(1 -\phi _{min})]\} \\
    e^{H_{min}} &=\exp \{ -[ \delta \alpha \tau _{maj}(1 - \phi _{maj}) + (1 -\delta) \alpha \tau _{min}\phi _{min}]\} \\
    e^{L_{maj}} &=\exp \{ -[ \delta (1 - \alpha) \tau _{maj} \phi _{maj} + (1 -\delta) (1 - \alpha) \tau _{min}(1 -\phi _{min})]\} \\
    e^{L_{min}} &=\exp \{ -[ \delta (1 - \alpha) \tau _{maj}(1 - \phi _{maj}) + (1 -\delta) (1 - \alpha) \tau _{min}\phi _{min}]\} 
\end{split}
\end{align*}
Then, the likelihood that a high-ability white worker accepts a referral wage, relative to the likelihood that a high-ability black worker accepts a referral wage can be calculated as follows:%
\begin{align*}
    \Pr\{H_{maj} \text{ accepts referral relative to }\Pr\{H_{min}\text{ accepts referral}\}\} = \gamma_{maj} =  \frac{1-e^{H_{maj}}}{1-e^{H_{min}}}
\end{align*}

\subsection{Step 2: Estimate the market wage and the expected referral wage for both racial groups}
\textit{Market wage}---As the proof for Proposition 3 in Appendix \ref{sec:appendixa} illustrates, Bayes's rule allows one to calculate the period-2 market wage:
\begin{align*}
\begin{split}
    & w_{M2}(\alpha, \delta , \tau _{maj}, \tau _{min}, \psi _{maj}, \psi _{min}) \\
    &= E(\text{productivity \textbar{} market}) \\
    &=\frac{\Pr(\text{market \textbar{} $H_{maj}$}) \cdot \Pr(H_{maj})+\Pr(\text{market \textbar{} $H_{min}$}) \cdot \Pr(H_{min})}{\Pr(\text{market \textbar{} $H$}) \cdot \Pr(H)+\Pr(\text{market \textbar{} $L$}) \cdot \Pr(L)} \\
    &=\frac{(e^{H_{maj}} \cdot \delta) +(e^{H_{min}} \cdot (1 -\delta))}{(e^{H_{maj}} + e^{L_{maj}}) \cdot \delta  +(e^{H_{min}} +e^{L_{min}}) \cdot (1 -\delta )}
\end{split}
\end{align*}

As the exposition for Proposition 2 mentions, $w_{M2}$ should always be below the average productivity of the workforce, which is 0.5.  This is because homophily along ability (that is, ability in-group bias) causes a disproportionately high volume of high-ability workers to be hired via referrals, which lowers the average productivity in the (non-referral) general market and drives down the equilibrium market wage.  

\renewcommand{\theequation}{B\arabic{equation}}
\setcounter{equation}{0}

\textit{Referral Wage}---Adapting Appendix Equation \ref{eq:maxreferral}, any given referral wage ${w_{R_k}}$ can be expressed as:%
\begin{align}
\begin{split}
    w_{R_k} =  & \frac{\begin{aligned}
        & p^{H_{maj}} \cdot \exp \{ -[p^{H_{maj}}][1 -F(w_{R_k})\}
        + p^{H_{min}} \cdot \exp \{ -[p^{H_{min}}][1 -F({w_{R_k}})]\} - c \end{aligned}%
        }{\begin{aligned}
            & p^{H_{maj}} \cdot \exp \{ -[p^{H_{maj}}][1 -F(w_{R_k})\}
            + p^{H_{min}} \cdot \exp \{ -[p^{H_{min}}][1 -F({w_{R_k}})]\} \\
            & +p^{L_{maj}} \cdot \exp \{ -[p^{L_{maj}}][1 -F({w_{R_k}})]\} 
            +p^{L_{min}} \cdot \exp \{ -[p^{L_{min}}][1 -F({w_{R_k}})]\}\end{aligned}}
\end{split}
\end{align}
where $F(\bullet)$ represents the equilibrium wage distribution; $c$ represents the expected firm profit, which is constant across the entire wage distribution and can thus be set equal to the expected productivity in the (non-referral) general market; and $p^X$ is denoted as follows:
\begin{align*}
\begin{split}
        p^{H_{maj}} &= \delta \alpha \tau _{maj} \phi _{maj} +(1 -\delta )\alpha \tau _{min}(1 -\phi _{min}) \\
        p^{H_{min}} &= \delta \alpha \tau _{maj} (1 - \phi _{maj}) + (1 -\delta) \alpha \tau _{min} \phi _{min} \\
        p^{L_{maj}} &= \delta (1 -\alpha) \tau _{maj} \phi _{maj} + (1 -\delta) (1 - \alpha) \tau _{min} (1 - \phi _{min}) \\
        p^{L_{min}} &= \delta (1 - \alpha) \tau _{maj} (1 - \phi _{maj}) + (1 -\delta) (1 - \alpha) \tau _{min} \phi _{min} \\\\
        c = E\prod _{H}(w_{M2}) &= (p^{HMAJ})(e^{HMAJ})(1 -w_{M2}) +(p^{HMIN})(e^{HMIN})(1 -w_{M2}) \\
        &\qquad  +( p^{LMAJ})(e^{LMAJ})(-w_{M2}) +(p^{LMIN})(e^{LMIN})(-w_{M2})
        \end{split}
\end{align*}
For simplicity, I assume a uniform equilibrium wage distribution $F(\bullet)$ and so have normalized the value for the average referral wage for black workers---$F(w_{R_k}^{H_{min}})$---to $0.5$ (since in Step 1 we normalized $\gamma_{min}=1$).  I then estimate $F(w_{R_k}^{H_{maj}})$ for the referral wage for white workers.  Since workers accept the maximum referral wage offered, estimating $F(w_{R_k}^{H_{maj}})$ is akin to finding the expected value of the maximum of $\gamma_{maj}$ random variables:\footnote{The PDF for a uniform distribution is $f_Y(y)=n \cdot y^{n-1}$.  This leads to $E[Y]= \int_{y\in A}yf_Y(y)dy=\int^1_0yny^{n-1}dy=\frac{n}{n+1}$.}
\begin{align*}
    F(w_{R_k}^{H_{maj}}) = \frac{\gamma_{maj}}{\gamma_{maj}+1}
\end{align*}

\subsection{Step 3: Estimate the expected welfare gap between racial groups}
To calculate the expected welfare gap between black and white workers, I take the sum of the market wage and the expected referral wage, weighted by the likelihood of being hired via the general market or via the referral market, respectively:
\begin{align*}
    \text{Welfare gap} & = 1-\frac{\Pr(H_{min}\text{ accepts }w_{M2})\cdot w_{M2} + \Pr(H_{min}\text{ accepts }w_R)\cdot E(w_{R_k}^{H_{min}})}
    {\Pr(H_{maj}\text{ accepts }w_{M2})\cdot w_{M2} + \Pr(H_{maj}\text{ accepts }w_R)\cdot E(w_{R_k}^{H_{maj}})} \\ 
    & = 1- \frac{e^{H_{min}} \cdot w_{M2} + (1-e^{H_{min}})\cdot E(w_{R_k}^{H_{min}})}
    {e^{H_{maj}} \cdot w_{M2} + (1-e^{H_{maj}})\cdot E(w_{R_k}^{H_{maj}})}
\end{align*}

\pagebreak

\printbibliography
\clearpage

\end{document}